\documentclass[twocolumn]{aa} 
\usepackage{graphicx}
\usepackage[varg]{txfonts}
\usepackage{natbib}
\usepackage[version-1-compatibility]{siunitx}
\DeclareSIUnit{\Msol}{\, \mathrm{M_\odot}}
\DeclareSIUnit{\Lsol}{\, \mathrm{L_\odot}}
\DeclareSIUnit{\Msolyr}{\, \mathrm{M_\odot \, yr^{-1}}}
\DeclareSIUnit{\ar}{\mathrm{a_R}}
\newcommand{\norm}[1]{\left\lVert#1\right\rVert} 
\newcommand{\ramses}{\textsc{Ramses}}

\newcommand{\hii}{H\textsc{ii}}

\newcommand{\be}{\begin{equation}}
\newcommand{\ee}{\end{equation}}
\newcommand{\beal}{\begin{aligned}}
\newcommand{\eeal}{\end{aligned}}
\newcommand{\mc}{M_\mathrm{c}}
\newcommand{\rc}{R_\mathrm{c}}
\newcommand{\rmc}{\mathrm{c}}
\newcommand{\rmg}{\mathrm{G}}

\usepackage{hyperref}
\bibpunct{(}{)}{;}{a}{}{,} 
\usepackage{amsmath}
\usepackage{esdiff}
\usepackage{xcolor}
\usepackage[english]{babel}

\begin{document}

   \title{Collapse of turbulent massive cores with ambipolar diffusion and hybrid radiative transfer \\
   I. Accretion and multiplicity}

   \author{R. Mignon-Risse
          \inst{1,2}
          \and
          M. Gonz\'alez \inst{1}
          \and
          B. Commer\c con \inst{3}
          \and
          J. Rosdahl \inst{4}
          }

   \institute{AIM, CEA, CNRS, Universit\'e Paris-Saclay, Universit\'e de Paris, F-91191 Gif-sur-Yvette, France \\
         \email{raphael.mignon-risse@apc.in2p3.fr}
         \and
         Universit\'e de Paris, CNRS, AstroParticule et Cosmologie, F-75013, Paris
         \and
              Univ Lyon, Ens de Lyon, Univ Lyon1, CNRS, Centre de Recherche Astrophysique de Lyon UMR5574, F-69007, Lyon, France
             \and
             Univ Lyon, Univ Lyon1, ENS de Lyon, CNRS, Centre de Recherche Astrophysique de Lyon UMR5574, F-69230, Saint-Genis-Laval, France
             }

   \date{Received ?; ?}
	\titlerunning{Collapse of turbulent massive cores with ambipolar diffusion and hybrid radiative transfer I.}
 
  \abstract
   {Massive stars form in magnetized and turbulent environments, and are often located in stellar clusters. The accretion and outflows mechanisms associated to forming massive stars, as well as the origin of their system's stellar multiplicity are poorly understood.
    }
   {We study the influence of both magnetic fields and turbulence on the accretion mechanism of massive protostars and their multiplicity. We also focus on disk formation as a pre-requisite for outflow launching.}
   {We present a series of four radiation-magnetohydrodynamical simulations of the collapse of a massive magnetized, turbulent core of $100\Msol$ with the adaptive-mesh-refinement code \ramses{}, including a hybrid radiative transfer method for stellar irradiation and ambipolar diffusion. 
   We vary the Mach and Alfv\'enic Mach numbers to probe sub- and superalfv\'enic turbulence as well as sub- and supersonic turbulence regimes.}
   {Subalfv\'enic turbulence leads to single stellar systems while superalfv\'enic turbulence leads to binary formation from disk fragmentation following spiral arm collision, with mass ratios of $1.1-1.6$ and a separation of several hundreds AU increasing with the initial turbulent support and with time. 
   In those runs, infalling gas reaches the individual disks via a transient circumbinary structure.
   Magnetically-regulated, thermally-dominated (plasma beta $\beta>1$), Keplerian disks form in all runs, with sizes $100-200$~AU and masses $1-8\Msol$. The disks around primary and secondary sink particles share similar properties.
   We obtain mass accretion rates of ${\sim}10^{-4} \Msolyr$ onto the protostars and observe higher accretion rates onto the secondary stars than onto their primary star companion. 
   The primary disk orientation is found to be set by the initial angular momentum carried by turbulence rather than by magnetic fields.
  Even without turbulence, axisymmetry and north-south symmetry with respect to the disk plane are broken by the interchange instability and the presence of thermally-dominated streamers, respectively.
   }
   {Small ($\lesssim 300$~AU) massive protostellar disks as those frequently observed nowadays can only be reproduced so far in the presence of (moderate) magnetic fields with ambipolar diffusion, even in a turbulent medium. The interplay between magnetic fields and turbulence sets the multiplicity of stellar clusters. A plasma beta $\beta>1$ is a good indicator to distinguish streamers and individual disks from their surroundings.}


   \maketitle
%
\section{Introduction}

Massive stars ($>8\Msol$) are luminous, among the main sources of feedback at parsec and galactic scales, especially due to their explosion into supernov\ae{}  \citep{larson_effects_1974}. Nonetheless, the conditions in which they form remain unclear.
Indeed, this challenging problem offers observational issues: massive stars are rare, located far away from us ($\gtrsim 1$~kpc) and in dense regions; and theoretical difficulties: in addition to magnetic fields and kinetic energy from turbulence \citep{tan_massive_2014}, radiative energy is to be accounted for.
At late stages, they are also expected to drive \hii{} region expansion \citep{keto_formation_2007}, bringing even more complexity to their environmental conditions of birth.
Hence, while the accretion mechanism in low-mass star formation is becoming increasingly understood in terms of disk-mediated accretion, it is not settled yet for high-mass star formation.
Furthermore, several outflow models rely on disk accretion, hence a first uncertainty on the origin of outflows follows from the uncertainty on the accretion process.
Finally, massive stars are located in stellar systems of higher multiplicity than low-mass stars \citep{duchene_stellar_2013}, but the origin of this trend is unknown, in particular whether it arises from core or disk fragmentation.
These three questions: accretion, ejection, and fragmentation (as a cause for multiplicity), are tightly linked together and depend on common physical ingredients: magnetic fields, turbulence, and radiation, that need to be modelled together.
All of them are addressed in this suite of two papers.

Low-mass star formation is better understood than high-mass star formation and first studies neglected the two last ingredients, namely turbulence and radiation, thus the first attempts to model massive star formation have built on this.
First of all, in the Competitive Accretion scenario \citep{bonnell_competitive_2001}, low- and high-mass protostars feed from a common gas reservoir after large-scale core fragmentation.
In this case, the final stellar mass is uncorrelated to the core mass.
Conversely, the scaled-up version of low-mass star formation, presented in the Turbulent Core accretion model \citep{mckee_formation_2003}, proposes that massive stars form in isolation from the collapse of high-mass pre-stellar cores, stabilized against gravitational collapse by turbulent motions and magnetic fields.
This model suffers from the rareness of high-mass pre-stellar cores \citep{motte_high-mass_2018}, though candidates do exist (\textit{e.g.}, \citealt{nony_detection_2018}), and may not be the most common procedure for massive star formation.
Global models, such as the Global Hierarchical Collapse model (GHC hereafter, \citealp{vazquez-semadeni_hierarchical_2016}) and the Inertial-Inflow model (II hereafter, \citealp{padoan_origin_2019}) challenge this core-fed accretion scenario and give prominence to large-scale dynamics, either due to collapse (GHC) or to the inertial motions following supernova feedback (II).
Those suggest the inclusion (and possible requirement) of turbulence.
It is not sure yet whether accretion should be "clump-fed" or "core-fed" and occur via disks or turbulent filaments (see \textit{e.g.}, \citealt{rosen_massive_2019}), but recent observations put (sparse but) increasingly convincing constraints on disk-mediated accretion, thanks to unprecedented angular resolution.


Current observational constraints on disks indicate radii of $20$ to $330$~AU (\citealt{patel_disk_2005}, \citealt{girart_resolving_2018}, \citealt{kraus_hot_2010}) and masses of $1-8\Msol$ \citep{patel_disk_2005}.
A complete massive star formation model should explain how these variations can be explained by the protostar's environment (magnetic field strength, turbulence, geometry?).
Disks can be subject to fragmentation, possibly leading in the formation of multiple stars gravitationally bound together.
The massive hot core region G351.77-0.54 observed with ALMA at sub-$40$~AU resolution reveals twelve sub-structures within a few thousand AU with a broad range of core separations \citep{beuther_high-mass_2019}, consistent with thermal Jeans fragmentation of a dense core, and possibly with the Global Hierarchical Model \citep{vazquez-semadeni_hierarchical_2016}. 
The aforementionned disk in HH 80-81 could be prone to fragmentation \citep{fernandez-lopez_rotating_2011} as well.
There does not seem to be any general trend about the disk stability, but the advent of ALMA will increase the statistics.
In the high-mass star-forming region IRAS 23033+5951, four mm-sources are identified with the Northern Extended Millimeter Array (NOEMA) and the IRAM telescope, 
out of which two exhibit protostellar activity.
Among those, one is stable and the other is prone to fragmentation in the inner $2000$~AU \citep{bosco_fragmentation_2019}.
Disk fragmentation may either lead to the formation of companion stars or to the accretion of clumps onto the central object.
Moreover, accretion leads to a radiative shock at the stellar surface and the energy is radiated away, hence clump accretion can be detected by luminosity outbursts.
The process behind this accretion luminosity is not fully understood, in particular the conditions under which the radiation would escape rather than being advected together with the gas.
It seems, however, a recurrent mechanism in low-mass star formation, and relies on disk-mediated accretion and star-disk interaction. Hence, it advocates the same accretion method for the formation of high-mass stars as well \citep{caratti_o_garatti_disk-mediated_2017}.

Altogether, despite the lack of systematic constraints, the presence of disks around young massive protostars ($L < 10^5 L_\odot$) is now well-established (see the reviews by \citealt{beltran_accretion_2016}, \citealt{beltran_disks_2020}).
Their properties may set the initial conditions for the formation of multiple stellar systems, and they strongly depend on the threading magnetic fields.


Constraints on magnetic field structures and strength are recent, due to new polarimetric instruments.
In a sample of 21 high-mass star-forming clumps, sub-parsec magnetic fields appear to be structured \citep{zhang_magnetic_2014}.
The hour-glass shape due to field lines being pulled by the collapsing gas is present \citep{beltran_alma_2019}, as in low-mass protostellar systems (\textit{e.g.} \citealt{maury_magnetically_2018}).
The parameter $\mu = (M/\phi)/(M/\phi)_\mathrm{crit}$ is the mass-to-flux to critical mass-to-flux ratio, where $\phi$ is the magnetic flux.
It indicates whether magnetic fields can ($\mu < 1$) or cannot ($\mu >1$) prevent collapse on their own.
Nevertheless, a $\mu \gtrsim 1$ still affects the gas dynamics.
Several studies agree on supercritical values of $\mu=1-4$ \citep{falgarone_cn_2008} or even $\mu \sim 1-2$ (\citealt{girart_magnetic_2009}, \citealt{li_self-similar_2015}, \citealt{pillai_magnetic_2015}), suggesting an important role of magnetic fields.
Quantitatively, the field strength has the order of $0.1-1$~mG in a sample of infrared dark clouds (IRDCs,  \citealp{pillai_cn_2016}), and in an ultracompact \hii{} region (UC\hii{}, \citep{tang_evolution_2009}), based on the Chandrasekhar-Fermi method\footnote{The Chandrasekhar-Fermi method relates the plane-of-the-sky field strength with the line-of-sight velocity dispersion, using the phase velocity of transverse Alfvén waves.} \citep{chandrasekhar_magnetic_1953}.
Comparisons with magneto-hydrodynamical simulations have shown that fragmentation is consistent with turbulence dominating over the magnetic energy (\citealt{palau_early_2013}, \citealt{fontani_magnetically_2016}).
Nonetheless, magnetic energy has been found to be comparable to (\citealt{falgarone_cn_2008}, \citealt{girart_dr_2013}) or to dominate over the turbulent energy (sub-alfv\'enic turbulence, \citealt{pillai_magnetic_2015}) in several sources.
\cite{girart_dr_2013} have found equipartition between rotational energy, magnetic energy and turbulent energy in a fast rotating core, with $\mu =6$, indicating three mechanisms capable of slowing down the collapse.

Hence, observations agree on the presence of disk-like structures, with several occurrences of fragmentation and sizes of tens to hundreds AU.
Magnetic fields have non-negligible strengths and their presence may affect both disk formation (and subsequently outflow launching), and its fragmentation into multiple stellar systems. 
Let us summarize the recent improvements made on disk formation on the side of numerical studies, focusing on the physics they have included and in particular the treatment of MHD and radiative transfer.

Disk-mediated accretion for massive protostars has emerged in multi-dimensional simulations as part of the so-called flashlight effect (\citealt{yorke_formation_2002}, \citealt{kuiper_circumventing_2010}), to overcome the radiation barrier problem \citep{larson_formation_1971}.
This effect describes the disk thermal radiation being radiated preferentially off the plane, ending in a small radiative force against the accretion flow, compared to the unidimensional view.
Meanwhile, progress has been made in the low-mass star formation context with the inclusion of magnetic fields in numerical simulation in the ideal magneto-hydrodynamics (MHD) frame (\textit{e.g.} \citealt{fromang_high_2006}).
Many studies have shown that in a collapsing core, the flux-freezing condition leads to the accumulation of magnetic fields in the central region, inducing a strong magnetic braking and preventing disk formation (see \textit{e.g.} \citealt{hennebelle_magnetic_2008}, and \citealt{seifried_magnetic_2011} in the high-mass regime).
This is referred to as the magnetic catastrophe, since many disks are observed around low- and high-mass protostars \citep{cesaroni_study_2005}.
Three ingredients have been introduced and shown separately to allow for the formation of disks and reconcile numerical simulations and observations in that respect: misalignment between the rotation axis and the magnetic field axis, turbulence, and non-ideal MHD effects.
Misalignment \citep{joos_protostellar_2012} and turbulence (\citealt{joos_influence_2013}, \citealt{lam_disk_2019} for low-mass stars, \citealt{seifried_disc_2012} for high-mass) directly reduce the magnetic braking efficiency.
Non-ideal (also called resistive) MHD effects, namely ambipolar diffusion (AD), Ohmic dissipation and the Hall effect provide a mechanism to limit the accumulation of magnetic fields strength and therefore the magnetic braking.
AD is probably the most-studied non-ideal MHD effect, as it starts dominating at lower densities than the others, and indeed promotes disk formation in the low- \citep{masson_ambipolar_2016} and high-mass regime (\citealt{commercon_discs_2021}, hereafter C21).
In several studies, non-ideal MHD appears as the main regulator of disk formation (AD in \citealt{hennebelle_magnetically_2016}), even when subsonic turbulence is included \citep{wurster_non-ideal_2020-1}.

In parallel, most numerical studies on massive star formation have focused on the radiative transfer aspect, due to the radiation pressure barrier \citep{larson_formation_1971}, and neglected magnetic fields.
First radiation-hydrodynamical implementations have relied on the Flux-Limited Diffusion (FLD) approximation \citep{levermore_flux-limited_1981} to describe infrared radiation interacting with dust-gas mixture.
The FLD method is well-suited for radiation transport in optically-thick media but is not adapted to strongly anisotropic radiation fields.
The particular treatment of the stellar radiation, also called irradiation, has been improved later on to track the higher-energy stellar photons and compute accurately the opacity during their first interaction with the ambient medium (\citealt{kuiper_fast_2010}, \citealt{flock_radiation_2013}, \citealt{ramsey_radiation_2015}, \citealt{rosen_hybrid_2017}, \citealt{mignon-risse_new_2020},\citealt{gressel_global_2020}, \citealt{fuksman_two-moment_2020}).

In the meantime, the common inclusion of radiative transfer and MHD in numerical codes has shown that both effects contribute to limit fragmentation.
\cite{commercon_collapse_2011} showed the prevention of early core fragmentation, while \cite{myers_fragmentation_2013} obtained similar results at later times.
Secondary fragmentation, responsible for the formation of companion stars, is also inhibited, as found by \cite{peters_interplay_2011}.

As presented above, the modelling of magnetized disks requires non-ideal MHD effects to circumvent the so-called magnetic catastrophe (C21).
In this work, we extend the C21 study and present the first numerical simulations including both a hybrid radiative transfer method and non-ideal MHD (namely, ambipolar diffusion), aiming at identifying the accretion conditions of massive protostars, as well as their outflow launching mechanism (Mignon-Risse et al, in prep., hereafter Paper II) with realistic physical ingredients.
To do so, we consider an initial velocity field consistent with turbulence (of various amplitudes, corresponding to several runs) in order to mimic non-idealized environmental conditions for the birth of a massive protostar.

This study is organized as follows.
The numerical methods are presented in Sect.~\ref{sec:model}.
In Sect.~\ref{sec:res} we analyze the disk-mediated accretion, emphasizing on the primary, secondary and circumbinary disks properties and on the disk-magnetic field alignment.
Our results and limitations are discussed in Sect.~\ref{sec:discussion} and we conclude in Sect.~\ref{sec:ccl}.

\section{Methods}
\label{sec:model}

In this section we present the set of equations that are solved numerically, and the set of initial conditions. We summarize our sink particle algorithm and finally, we present the physical criteria that define a disk in the post-processing step.
This work is intended to extend the study of C21 that has been done with the Flux-Limited Diffusion method for radiative transfer, but this time with an hybrid radiative transfer method and in a turbulent medium. An additional difference resides in the sublimation model we take here, and an optically-thin sink volume (see below).

\subsection{Radiation magneto-hydrodynamical model}

We integrate the equations of radiation-magneto-hydrodynamics (MHD) in the adaptive-mesh refinement (AMR) \ramses{} code (\citealt{teyssier_cosmological_2002}, \citealt{fromang_high_2006}) with ambipolar diffusion \citep{masson_incorporating_2012},
and the so-called hybrid radiative transfer method \citep{mignon-risse_new_2020}, namely the M1 method (\citealt{levermore_relating_1984}, \citealt{rosdahl_ramses-rt:_2013}, \citealt{rosdahl_scheme_2015}) for stellar radiation and the Flux-Limited Diffusion (FLD, \citealt{levermore_flux-limited_1981}, \citealt{commercon_radiation_2011}, \citealt{commercon_fast_2014}) otherwise.
The set of equations we solve is
   \begin{equation}
   \begin{aligned}
   \diffp{\rho}{t} + \nabla \cdot [\rho \boldsymbol{u}] 
   &= 0, \\
   \diffp{\rho \boldsymbol{u}}{t} + \nabla \cdot [\rho \boldsymbol{u} \otimes \boldsymbol{u} + P \mathbb{I}]
   &= - \lambda \nabla E_\mathrm{fld} + \frac{\kappa_\mathrm{P,\star} \rho}{\mathrm{c}} \boldsymbol{F}_\mathrm{M1} +\boldsymbol{F}_\mathrm{L} - \rho \nabla \phi, \\
   \diffp{E_\mathrm{T}}{t} + \nabla \cdot \biggl[\boldsymbol{u} \left( E_\mathrm{T} + P + B^2/2 \right) \\
   - (\boldsymbol{u} \cdot \boldsymbol{B}) \boldsymbol{B} - \boldsymbol{E}_\mathrm{AD} \times \boldsymbol{B} \biggr]
   &= - \mathbb{P}_\mathrm{fld} \nabla : \boldsymbol{u} + \kappa_\mathrm{P,\star} \, \rho \mathrm{c} E_\mathrm{M1}  - \lambda \boldsymbol{u} \nabla E_\mathrm{r} \\
   & + \nabla \cdot \left( \frac{\mathrm{c} \lambda}{\rho \kappa_{\mathrm{R,fld}}}\nabla E_\mathrm{r} \right) - \rho \boldsymbol{u} \cdot \nabla \phi, \\
   \diffp{E_{\mathrm{M1}}}{t}  + \nabla \cdot \boldsymbol{F}_\mathrm{M1}
   &= - \kappa_\mathrm{P,\star} \, \rho \mathrm{c} E_\mathrm{M1} + \dot{E}_\mathrm{M1}^\star, \\
   \diffp{\boldsymbol{F}_\mathrm{M1}}{t}  + \mathrm{c}^2 \nabla \cdot \mathbb{P}_\mathrm{M1}
   &= - \kappa_\mathrm{P,\star} \, \rho \mathrm{c} \boldsymbol{F}_\mathrm{M1}, \\
  \diffp{E_{\mathrm{fld}}}{t} 
   - \nabla \cdot \left( \frac{\mathrm{c} \lambda}
   	{\rho \kappa_{\mathrm{R,fld}}} \nabla E_{\mathrm{fld}} \right)
   &= \kappa_{\mathrm{P,fld}} \, \rho \mathrm{c} \left( \ar T^4 - E_{\mathrm{fld}} \right), \\
   \diffp{\boldsymbol{B}}{t} - \nabla \times \left[ \boldsymbol{u} \times \boldsymbol{B} + \boldsymbol{E}_\mathrm{AD} \right]
   &= 0, \\
   \nabla \cdot \boldsymbol{B} &= 0, \\
   \Delta \phi &= 4 \pi \mathrm{G} \rho.
   \end{aligned}
   \end{equation}
   Here, $\rho$ is the material density, $\boldsymbol{u}$ is the velocity, $P$ is the thermal pressure, $\lambda$ is the FLD flux-limiter, $E_\mathrm{fld}$ is the FLD radiative energy, $\kappa_\mathrm{P,\star}$ is the Planck mean opacity at the stellar temperature, $\boldsymbol{F}_\mathrm{M1}$ is the M1 radiative flux, $\boldsymbol{F}_\mathrm{L} = (\nabla \times \boldsymbol{B}) \times \boldsymbol{B}$ is the Lorentz force, $\phi$ is the gravitational potential, $E_\mathrm{T}$ is the total energy $E_\mathrm{T} = \rho \epsilon + 1/2 \rho u^2 + 1/2 B^2 + E_\mathrm{fld}$ ($\epsilon$ is the specific internal energy), $E_\mathrm{M1}$ is the M1 radiative energy, $\boldsymbol{B}$ is the magnetic field, $\boldsymbol{E}_\mathrm{AD}$ is the ambipolar electromotive force, $\mathbb{P}_\mathrm{fld}$ is the FLD radiative pressure, $\kappa_\mathrm{P,fld}$ is the Planck mean opacity in the FLD module, $\kappa_\mathrm{R,fld}$ is the Rosseland mean opacity, $\ar$ is the radiation constant, $\mathbb{P}_\mathrm{M1}$ is the M1 radiative pressure, $\dot{E}_\mathrm{M1}^\star$ is the stellar radiation injection term.
   
   Let us note that we inject radiative energy into the M1 module only with the primary sink, while other sinks (formed after the primary one) will radiate within the FLD module. The justification is twofold. 
   First, in the M1 method the radiative fluxes from several sources sum up, whereas two radiation beams should not interact when crossing each other \citep{gonzalez_heracles:_2007}. The generalization of the hybrid method, which relies on the M1 method, to several stellar sources should be addressed in dedicated studies.
   Second, we address the origin of outflows around individual massive protostars (paper II), and using the M1 method for treating one star's radiation is sufficient to do so. 
   
   The ambipolar electromotive force is equal to
   \be
   \mathbf{E}_\mathrm{AD} =
\frac{\eta_\mathrm{AD}}{B^2} \left[  (\nabla \times \mathbf{B} )\times \mathbf{B} \right] \times \mathbf{B},
   \ee
   where $\eta_\mathrm{AD}$ is the ambipolar diffusion resistivity.
   It depends on the density, temperature, and magnetic field strength.
   The resistivities are pre-computed using a chemical network to calculate the equilibrium abundances of the molecules (neutrals) and main charge carriers in conditions of pre-stellar core collapse \citep{marchand_chemical_2016}, depending on the density and temperature.
   Chemical equilibrium is assumed because the associated timescale is shorter than the free-fall time at the densities we consider. 
   
   The term $\kappa_\mathrm{P,\star}\rho  \mathrm{c} E_\mathrm{M1}$ couples the M1 and the FLD methods via the equation of evolution of the internal energy
   \begin{equation}
    C_\mathrm{v} \diffp{T}{t} 
   = \kappa_\mathrm{P,\star} \, \rho \mathrm{c} E_\mathrm{M1}
   + \kappa_{\mathrm{P,fld}} \, \rho \mathrm{c} \left(E_{\mathrm{fld}} - \ar T^4  \right).
   \end{equation}
   We use the ideal gas relation for the internal specific energy $\rho \epsilon = C_\mathrm{v} T$
   where $C_\mathrm{v}$ is the heat capacity at constant volume.

\subsection{Physical setup}

We take similar initial conditions as C21.
The free-fall time in the central plateau (which contains ${\sim}15\Msol$) of the density profile is then
\begin{equation}
    \tau_\mathrm{ff} = \sqrt{\frac{3 \pi}{32 \mathrm{G} \Bar{\rho_\mathrm{0}}}} \simeq 24 \mathrm{\, kyr}, 
    \label{eq:tauff}
\end{equation}
where $\mathrm{G}$ is the gravitational constant and $\Bar{\rho_\mathrm{0}}$ is the density of the central plateau.

The initial core is threaded by a uniform magnetic field oriented along the $x-$axis.
We set the magnetic field strength by the mass-to-flux to critical mass-to-flux ratio $\mu = (M/\Phi)_\mathrm{0}/(M/\Phi)_\mathrm{crit}$ where $(M/\Phi)_\mathrm{0} = M_\mathrm{c}/(\pi B_\mathrm{0} R_\mathrm{c}^2)$ and $(M/\Phi)_\mathrm{crit} = 0.53/(3 \pi) \sqrt{5/G}$ \citep{mouschovias_note_1976}.
Strong ($\mu=2$, $B_\mathrm{0}=170\, \mu G$) and moderate ($\mu=5$, $B_\mathrm{0}=68\, \mu G$) magnetic fields are considered here.
A drawback of this uniform distribution is that the mass-to-flux ratio decreases as $\mu \, {\sim} \, 1/R$ and is larger in the inner parts of the core, with $\mu\, {\approx}\, 50$ in the central plateau (for runs with $\mu=5$) corresponding to a weakly magnetized medium.
We expect, however, the central magnetic field strength to increase as $B\varpropto \rho^{2/3}$ as the core contracts and before ambipolar diffusion starts dominating (at $\rho{\sim}10^{-15} \mathrm{g\, cm^{-3}}$). 
Thus, the magnetic field will play a dynamical role in the collapse (see C21).

An initial velocity dispersion is imposed to mimic a turbulent medium, and follows a Kolmogorov power spectrum $P(k) \varpropto k^{-5/3}$, similar to \cite{commercon_collapse_2011}. 
Phases are randomly sampled, and one realization is considered, \textit{i.e.} we do not vary the velocity field but only its amplitude between two runs. 
The turbulence is not sustained but the turbulent crossing time (${\sim} 0.5-2 \mathrm{\, Myr}$ with $T=20$~K, see below) is significantly larger than the simulation time here, so it should not impact our results. 
A low level of (solid-body) rotation, $E_\mathrm{rot}/E_\mathrm{grav}=1\%$, is initially imposed around the $x-$axis, and dominates the specific angular momentum in subsonic runs.
We consider four runs (see Table~\ref{table:tests}), varying the initial Mach number $\mathcal{M}$ and Alfvénic Mach number $\mathcal{M}_\mathrm{A}$.
The runs fall into two classes, that we will refer to throughout the paper, depending on the relative impact of turbulence and magnetic fields. Indeed, turbulence is subalfv\'enic in runs \textsc{NoTurb} ($\mathcal{M}_\mathrm{A} = 0$) and \textsc{SubA} ($\mathcal{M}_\mathrm{A} < 1$), and superalfv\'enic in runs \textsc{SupA} and \textsc{SupAS} ($\mathcal{M}_\mathrm{A} > 1$).
Regarding the Mach number, runs \textsc{SupA} and \textsc{SubA} have subsonic turbulence with $\mathcal{M} = 0.5$ while run \textsc{SupAS} has a supersonic turbulence with $\mathcal{M} = 2$.

Sink particles are introduced to mimic the presence of a protostar and radiate as blackbodies.
Their radii and internal luminosities (which give their effective surface temperature) are interpolated from a Pre-Main Sequence track \citep{kuiper_simultaneous_2013}, provided their mass at a given time and their accretion rate averaged over their lifetime. 
We model the dust sublimation by decreasing progressively the dust-to-gas ratio with the temperature. Gray opacities are taken from \cite{semenov_rosseland_2003} and the dust sublimation is modelled by a dust-to-gas ratio which vanishes at high temperatures, similarly to \cite{kuiper_circumventing_2010}.
The dust-to-gas ratio varies as
\begin{equation}
    \frac{M_\mathrm{dust}}{M_\mathrm{gas}} (\rho, T) =\left( \frac{M_\mathrm{dust}}{M_\mathrm{gas}} \right)_0 \left( 0.5 - \frac{1}{\pi} \mathrm{arctan} \left( \frac{T-T_\mathrm{evap}(\rho)}{100} \right) \right)
\end{equation}
where $\left( \frac{M_\mathrm{dust}}{M_\mathrm{gas}} \right)_0=1 \%$ is the initial dust-to-gas mass ratio,
and the evaporation temperature is
\begin{equation}
    T_\mathrm{evap}(\rho) = g \left( \frac{\rho}{1 \mathrm{\, g \, cm^{-3}}} \right)^\beta
\end{equation}
where $g = 2000\mathrm{\, K}$ and $\beta = 0.0195$ \citep{isella_shape_2005}.
The gas opacity is taken equal to $0.01 \mathrm{\, cm^2 \,g^{-1}}$, so the total opacity tends towards this value as the temperature increases beyond $T_\mathrm{evap}$.
Finally, we set the opacity in the primary sink particle volume to a value chosen so that the local optical depth is the minimal optical depth allowed numerically
($10^{-4}$).
This floor value for the optical depth is a numerical parameter used in optically-thin cells in order to gain performance with the FLD solver (see Appendix A of \citealt{vaytet_protostellar_2018}).
Gas and radiation are hardly modelled within the sink volume, but stellar radiation is meant to escape this volume.
This justifies our subgrid model for decoupling gas and radiation within the sink volume (see the discussion in Appendix~\ref{sec:app_fldhy}). 

\subsection{Resolution and sink particles}
\label{sec:resol}

Boundary conditions are periodic and the simulation box is $0.8\mathrm{\, pc}$ large, hence the gravitational effects due to the periodicity are marginal\footnote{The box length is twice the core diameter. At the core border, the gravitational acceleration exerted on a gas particle scales as $a_\mathrm{grav} {\sim}M_\mathrm{c}/R_\mathrm{c}^2$. In comparison, the acceleration due to the nearest ($0.6\mathrm{\, pc}$) periodic core is ${\sim} M_\mathrm{c}/(3R_\mathrm{c})^2 = a_\mathrm{grav}/9$.}. 
A cell effective resolution is $0.5^\ell$ (in units of the box length), where $\ell$ is the AMR level of refinement.
The coarse grid is $32^3$ (level $\ell=5$), with $10$ additional levels of refinement. 
It leads to a smallest cell width of $5$~AU. 
We run a similar set of simulations with a maximal resolution of $10$~AU to probe resolution convergence.
We use the prefix LR, as "low-resolution", to refer to these runs (not shown here for conciseness).
Refinement is performed on the Jeans length: it must be sampled by at least $12$ cells, following \cite{truelove_jeans_1997}.
Sinks are introduced at the finest AMR level, after a clump has been found to be bound and Jeans unstable (see C21, \citealt{bleuler_towards_2014}).
Sinks accrete the environmental material that enters their accretion volume, as follows. The radius of this sphere, so-called accretion radius is set to four times the finest resolution, \textit{i.e.} $20$~AU ($40$~AU for the LR runs).
We use a threshold accretion scheme based on the Jeans density in order to avoid artificial fragmentation.
If a sink cell has a density above the local Jeans density, $10\%$ of the excess is accreted by the sink particle within one time step.
All of the accreted mass goes into the sink particle mass (no outflow subgrid model).
Sink particles are collisionless, they interact only gravitationally with the gas and with their companions.
Their gravitational potential is prevented from diverging by the use of a so-called softening length, that is set equal to the accretion radius.
Finally, sink particles can merge if their accretion radii overlap.
We do not add any criterion to prevent sinks from merging. 

\begin{figure*}
\centering
    \includegraphics[width=18cm]{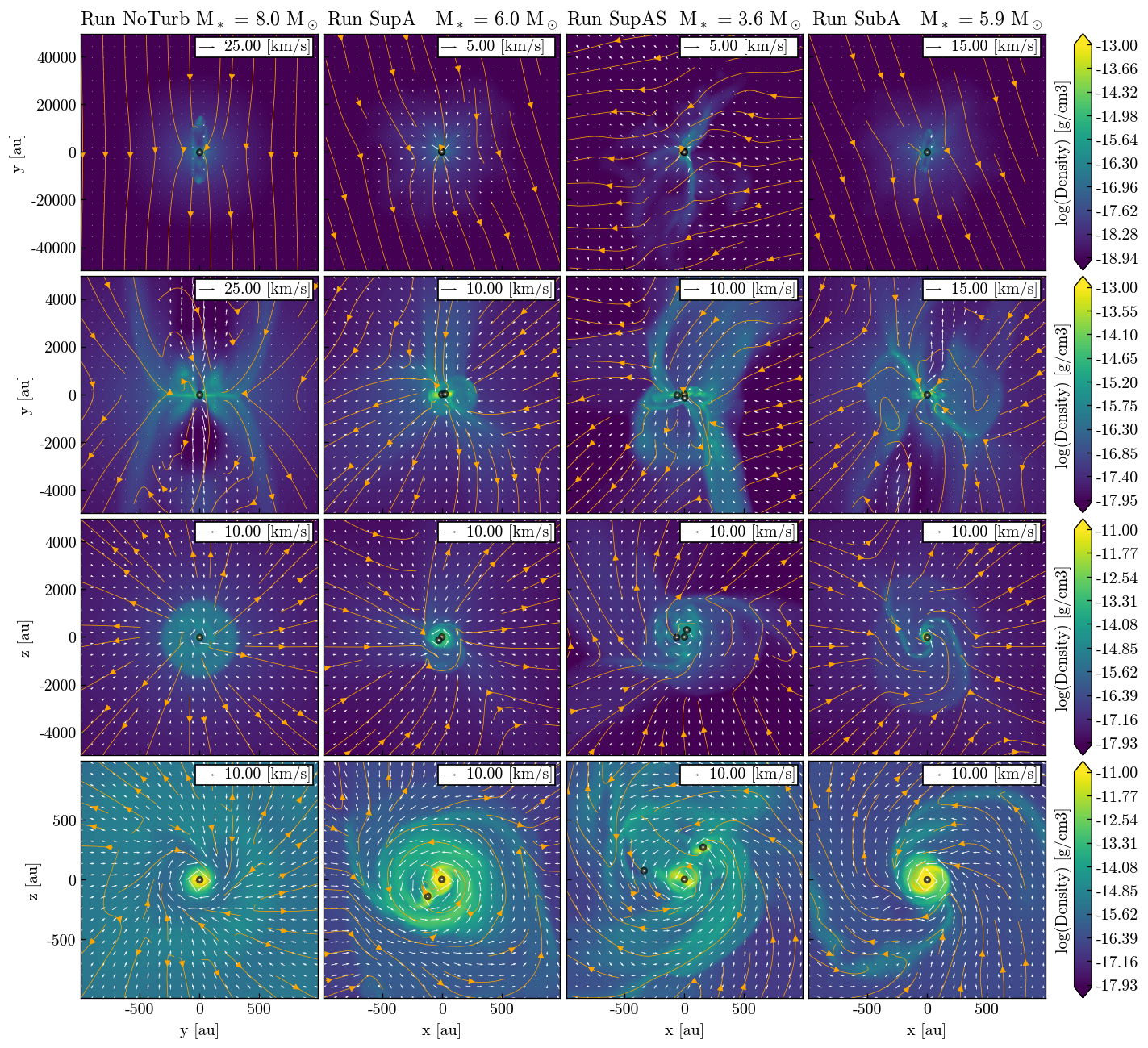}  
    \caption{Density slices perpendicular (first and second row, x10 zoom) and parallel (third and fourth row, x5 zoom) to the disk plane, at $t{\approx}50$~kyr. Streamlines corresponding to magnetic field lines, and arrows corresponding to the velocity field, are overplotted. Columns from left to right: run \textsc{NoTurb}, \textsc{SupA}, \textsc{SupAS}, \textsc{SubA}. A mass density $\rho=10^{-19} \mathrm{\, g\, cm^{-3}} $ corresponds to a particle density $n=2.6 \times 10^4 \mathrm{\, cm^{-3}}$ and $\rho=10^{-11} \mathrm{\, g\, cm^{-3}}$ to $n=2.6 \times 10^{12} \mathrm{\, cm^{-3}}$. White dots represent sink particles.}
    \label{fig:rhomaps}
\end{figure*}

\subsection{Disk identification}
\label{sec:diskdef}

Primary disk properties are presented in Sect.~\ref{sec:diskppt}.
These are computed from the cell-by-cell disk selection, which relies on the criteria presented in \cite{joos_protostellar_2012}
\begin{itemize}
    \item It is rotationally-supported: $\rho v_\phi^2/2 > f_\mathrm{thres} P$, where $v_\phi$ is the azimuthal velocity and $P$ is the thermal pressure. We choose $f_\mathrm{thres}=2$ as in \cite{joos_protostellar_2012};
    \item The gas number density is higher than $n=10^9 \mathrm{cm^{-3}}$, \textit{i.e.} $\rho=3.85 \times 10^{-15} \mathrm{g\, cm^{-3}}$;
    \item The gas is not about to free-fall onto the central object too rapidly: $v_\phi > f_\mathrm{thres} v_r$, where $v_r$ is the radial velocity;
    \item The disk vertical structure is in hydrostatic equilibrium: $v_\phi > f_\mathrm{thres} v_z$, where $v_z$ is the vertical velocity.
\end{itemize}
Let us note that there is no connectivity criterion, hence the extremity of a high-density spiral arm can be considered as part of the disk while the inter-arm low-density region (due to the gas being swept) may not be.
We compute the disk radius as the radius enclosing $90\%$ of the total disk selection mass, to avoid accounting for transient negligible contributions from larger scales.
We add a geometrical criterion to avoid perturbations from companions and their possible disks: the cell must be located less than $0.9$ times the binary separation. This is justified a posteriori as individual disks are observed around each star.

\begin{table}
\caption{Initial conditions of the four runs: name, Mach number, Alfv\'enic Mach number, mass-to-flux to critical mass-to-flux ratio, turbulence relative strength (with respect to thermal support and magnetic fields), respectively.}
\label{table:tests}
\centering 
\begin{tabular}{c | c c c | c} 
	\hline\hline
	 Model & $\mathcal{M}$ & $\mathcal{M}_\mathrm{A}$ &  $\mu$ & Turbulence relative strength \\ \hline \hline
	\textsc{NoTurb}   	&  0	 & 	0	  &  5 & No turbulence \\ \hline 
    \textsc{SupA}	 &  0.5   & 1.4		 & 5 & Superalfv\'enic, subsonic \\ \hline
	\textsc{SupAS}		 & 	2 &	 5.7  & 5 & Superalfv\'enic, supersonic\\  \hline
	\textsc{SubA}    &  0.5    &   0.57   & 2	& Subalfv\'enic, subsonic  \\ \hline
\end{tabular}
\end{table}

\begin{figure*}
\centering
    \includegraphics[width=18cm]{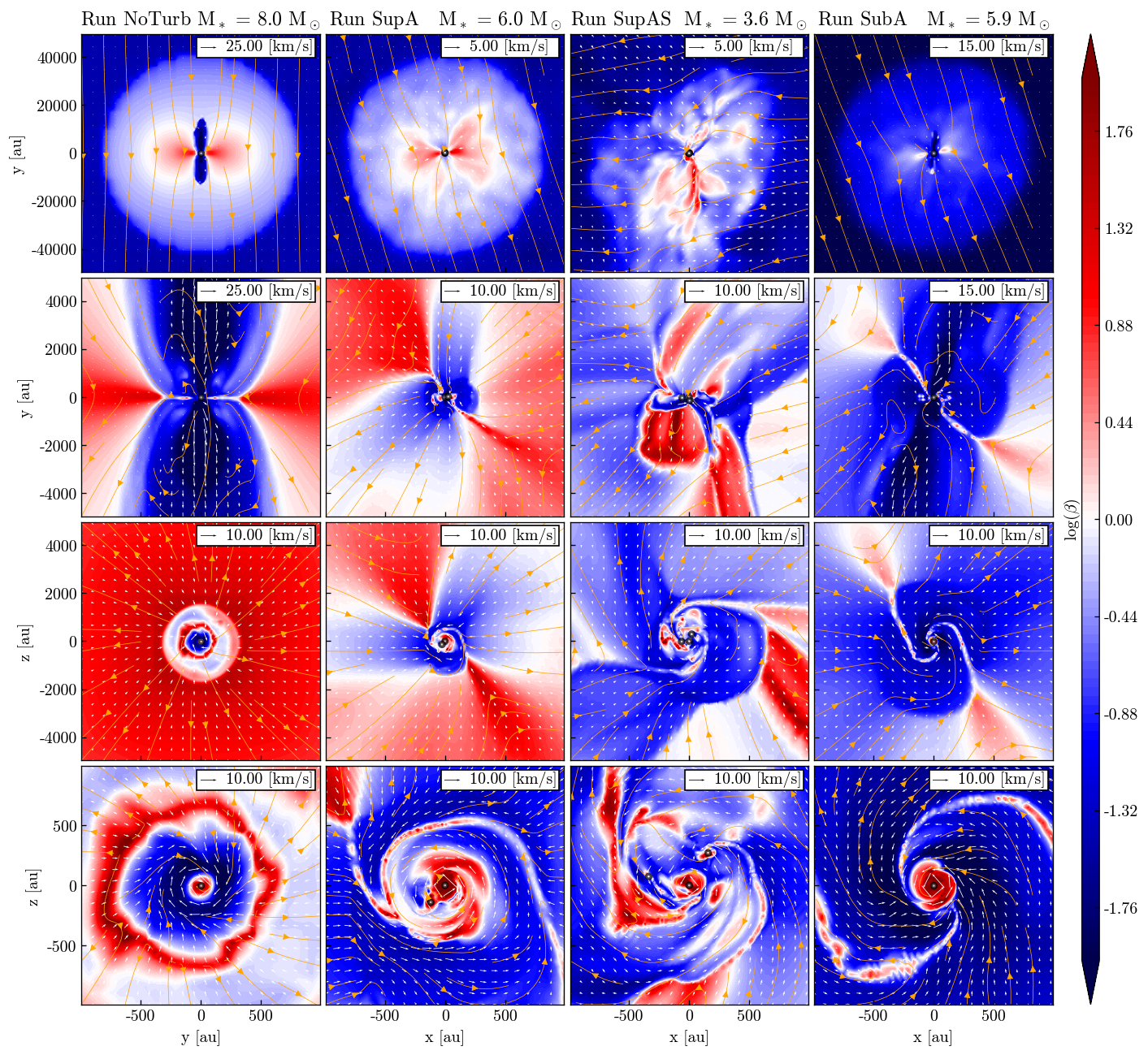}  
    \caption{Same as Fig.~\ref{fig:rhomaps} but with plasma beta (ratio between the thermal pressure and the magnetic pressure) slices. The gas moves along magnetic field lines until it forms thermally-dominated infalling filaments. Disks have $\beta>1$ too.}
    \label{fig:betamaps}
\end{figure*}

\section{Results}
\label{sec:res}

We first summarize some of the main and common features or all runs in Sect.~\ref{sec:overview}.
The sink mass evolution, which depends on disk-mediated accretion, is presented in Sect.~\ref{sec:sinkmass}.
Then, we focus on the properties of the primary disk (Sect.~\ref{sec:diskppt}), the secondary disk (Sect.~\ref{sec:seconddisk}) and the circumbinary disk (Sect.~\ref{sec:circumd}).
Finally, we study the primary disk alignment with the core-scale magnetic fields (Sect.~\ref{sec:alignangmom}). 

\subsection{Overview and common features}
\label{sec:overview}

\subsubsection{Formation of structures and stars}

In the four simulations, as the gravitationally-unstable cloud core collapses the first sink particles form at $t\,{\approx}\,29$~kyr.
Figure~\ref{fig:rhomaps} shows density slices in the disk plane and perpendicularly to the disk plane in the four runs, at time $t=50$~kyr, together with gas velocity and magnetic field lines.
Except in run \textsc{SupAS} (the most turbulent, third column of Fig.~\ref{fig:rhomaps}), where we get a large filament-like structure due to the stronger inner turbulent support (see below), the dense region ($\rho \gtrsim 10^{-16}\mathrm{g \, cm^{-3}}$) becomes rapidly concentrated in a sphere of diameter ${\sim}2000$~AU (center panels, second row of Fig.~\ref{fig:rhomaps}).
This is reminiscent of the structure described by \cite{machida_failed_2020} and attributed to the toroidal magnetic pressure not being able to launch an outflow because of turbulence.
Indeed, we show in Fig.~\ref{fig:betamaps} the plasma beta ($\beta=P_\mathrm{th}/P_\mathrm{mag}$, where $P_\mathrm{th}$ and $P_\mathrm{mag}$ are the thermal pressure and magnetic pressure, respectively) corresponding to Fig.~\ref{fig:rhomaps} and see that the central region in run \textsc{SupAS} (third column, second row) is magnetically-dominated ($\beta < 1$) on $2000$~AU scales while thermally-dominated ($\beta > 1$) matter is infalling.
This illustrates the importance to accurately account for the coupling between gas and magnetic fields in order to assess the dynamical role expected to be played by magnetic effects.
Accretion disks form in all runs around the primary sink (third and last rows of Fig.~\ref{fig:rhomaps}). 
In runs \textsc{NoTurb} and \textsc{SubA}, in which turbulence is subalfv\'enic, no secondary sink forms.
With superalfv\'enic turbulence (runs \textsc{SupA} and \textsc{SupAS}), a secondary long-lived sink particle forms in the primary sink accretion disk.
We will study in more details the stellar multiplicity in Sect.~\ref{sec:sinkmass}, and the secondary and circumbinary disks properties in Sect.~\ref{sec:seconddisk} and Sect.~\ref{sec:circumd}, respectively.
When not mentioned otherwise, we will refer to the primary sink and to the primary disk, for conciseness.

\subsubsection{Magnetic field evolution}

\begin{figure*}
\centering
    \includegraphics[width=7cm]{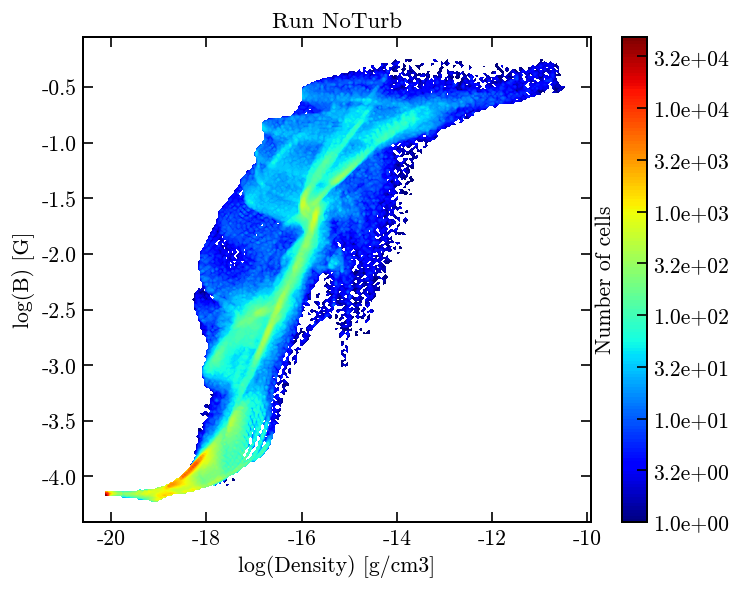}
    \includegraphics[width=7cm]{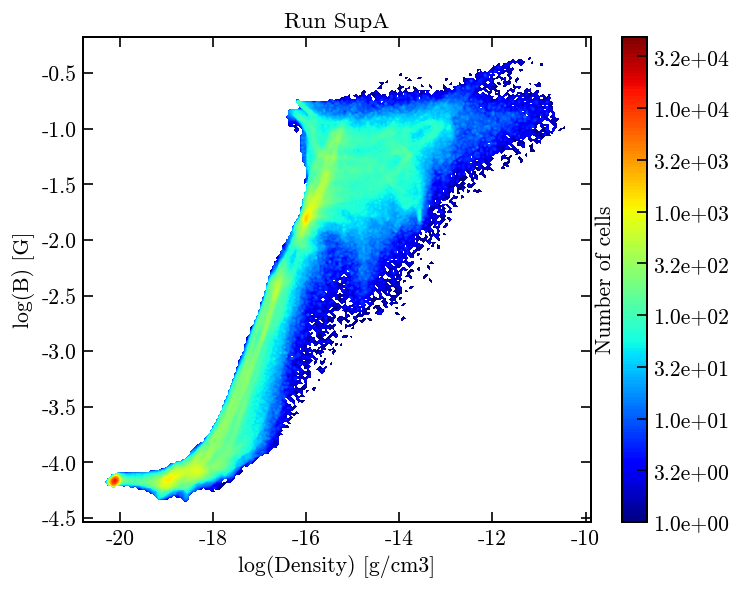}\\
    \includegraphics[width=7cm]{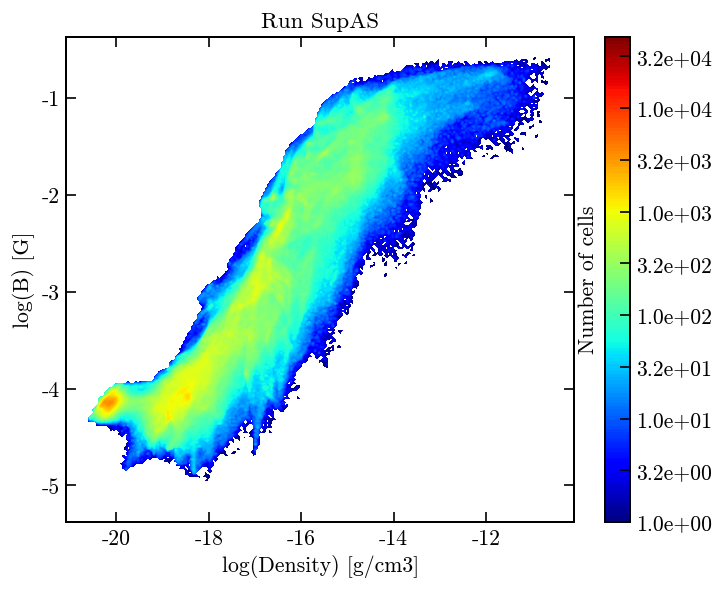}
    \includegraphics[width=7cm]{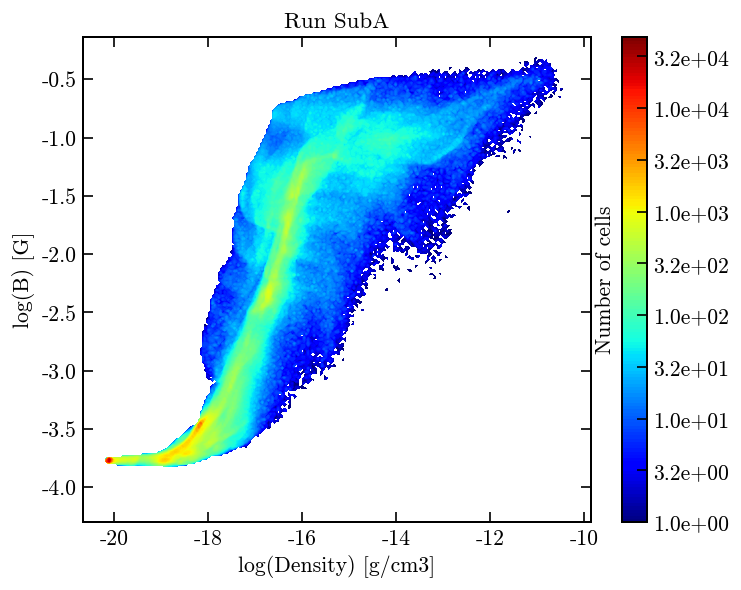}
    \caption{Density-magnetic field strength histograms at $t=50$~kyr. From left to right: run \textsc{NoTurb} and \textsc{SupA} (top), \textsc{SupAS} and \textsc{SubA} (bottom).}
    \label{fig:histograms}
\end{figure*}

As collapse occurs, the magnetic field strength is expected to increase in the central regions.
The density-magnetic field strength histograms for runs (from left to right) \textsc{NoTurb}, \textsc{SupA}, \textsc{SupAS} and \textsc{SubA} at $t=50$~kyr are shown in Fig.~\ref{fig:histograms}.
At densities below ${\sim}10^{-15} \mathrm{\, g\, cm^{-3}}$, we recover the ideal MHD limit where $B$ increases with $\rho$.
In runs \textsc{NoTurb} and \textsc{SubA}, the high-B, low-$\rho$ part of the histogram is populated by outflowing material ejected from the most magnetized regions.
At high densities, the plateau-like feature is present in the four runs, and contrasts with ideal MHD calculations, as shown in the low-mass  \citep{masson_ambipolar_2016} and high-mass (C21) regimes.
This is due to ambipolar diffusion, which becomes dominant above $\rho \gtrsim 10^{-15} \mathrm{g\, cm^{-3}}$.
The diffusion coefficient varies non-linearly with the magnetic field strength $\eta_\mathrm{AD}\varpropto B^2/\rho$ which explains its strong regulating effect.
The plateau is located between ${\sim}0.1$~G in the superalfv\'enic runs (\textsc{SupA}, \textsc{SupAS}), and ${\sim}0.3$~G in the subalfv\'enic runs (\textsc{NoTurb}, \textsc{SubA}).
The inclusion of ambipolar diffusion prevents the magnetic field strength to increase, which would change the disk structure and possibly the outflows since a strong magnetic field is reasonably expected in the magneto-centrifugal mechanism (C21).
The large dispersion observed in the bottom-left panel of Fig.~\ref{fig:histograms} at low density is provoked by turbulence in the core.

\subsubsection{Asymmetries}

While the numerical setup in run \textsc{NoTurb} is initially axisymmetric and symmetric with respect to the $(x=0,y,z)-$plane (that we will refer to as the "north-south" symmetry), these symmetries are broken.
First, we observe that pockets of magnetized plasma ($\beta<1$) are regularly expelled from the disk outer edge (top panels of Fig.~\ref{fig:beta_interch}).
This is visible in run \textsc{NoTurb} but hardly seen in the other, turbulent runs.
We investigate in Appendix~\ref{sec:app:intinstab} whether the magnetic interchange instability, which has been found to redistribute magnetic flux after accumulation around sink particles \citep{krasnopolsky_protostellar_2012} or at the ambipolar diffusion radius \citep{li_non-ideal_2011}, is responsible for this.
We find that the timescale associated with the interchange instability is indeed small enough (compared to the local advection timescale) to justify the interchange instability as a good candidate.
This phenomenon is not the only asymmetry arising in the simulation.
Indeed, we observe the presence of filamentary structures linking the densest regions (where the sink-disk system is) to the envelope, that we will refer to as "streamers".
These are visible in Fig.~\ref{fig:betamaps}, as the filaments that are dominated by thermal pressure ($\beta>1$), unlike the gas that surrounds them.
They have a density $\rho \gtrsim 10^{-15} \mathrm{g \, cm^{-3}}$.
They appear as a path for the accretion flow and pull the magnetic field lines along with them, which in turn form an hour-glass shape.
Since the Lorentz force has no component parallel to the field lines, the gas can move along the lines to join the streamers without any magnetic resistance, in a similar way as the bead-on-a-wire picture for magneto-centrifugal jets.
These streamers form perpendicularly to the core-scale magnetic field, in all runs.
In run \textsc{NoTurb}, this plane is also that of the accretion disk.
Nonetheless, they connect to the disk outside of the disk plane (first column, third and last rows of Fig.~\ref{fig:betamaps}), either from above or below, breaking the north-south symmetry.
We attribute this to the strong magnetic forces around the streamers.
In runs \textsc{SupA} and \textsc{SubA}, the streamers are much thicker.
This could be a hint of turbulent diffusion but it is not investigated further here. 
This gives rise to the filament-like structure of width ${\sim}2000$~AU in run \textsc{SupAS}, as mentioned above.
Overall, the streamers do not seem to set the disk formation plane (studied in Sect.~\ref{sec:alignangmom}).
Nevertheless, the symmetry breaking they provide is important to us, considering that $16\%$ of the outflows reported in \cite{wu_study_2004} are monopolar and our aim is to study a non-idealized case that could be compared to observations.


\begin{table}
\caption{Simulations outcomes: $M_{\star,\mathrm{end}}$ is primary star mass (in units of $\Msol$) at the time $t_\mathrm{end}$ (in kyr) of the run and $M_{2, \mathrm{end}}$ is the secondary sink mass at the same time.} 
\label{table:results}
\centering 
\begin{tabular}{c | c c c} 
    \hline \hline
	 Model & $t_\mathrm{end}$ & $M_{\star,\mathrm{end}}$ & $M_{2,\mathrm{end}}$ \\ \hline \hline
	\textsc{NoTurb}    & $80.4$ & $15.8$&  - \\ \hline 
    \textsc{SupA}	  & $71.5$ & $8.8$ & $7.7$ \\ \hline
	\textsc{SupAS}	  & $78.5$ & $6.2$ & $9.9$  \\ \hline
	\textsc{SubA} & $66.2$ & $11.0$ &  - \\ \hline
\end{tabular}
\end{table}

\subsubsection{Angular momentum transport}

\begin{figure*}
\centering
    \includegraphics[width=6cm]{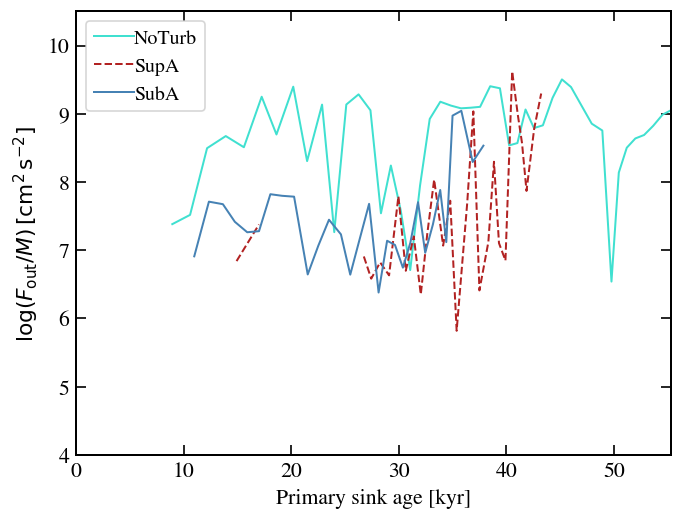}
    \includegraphics[width=6cm]{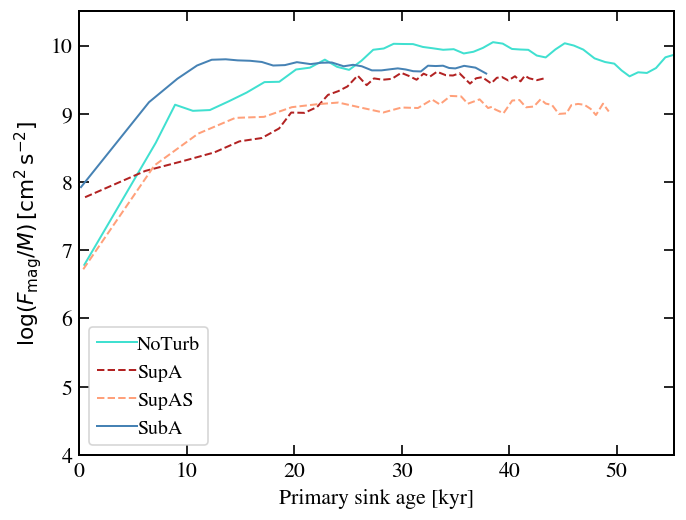}
    \includegraphics[width=6cm]{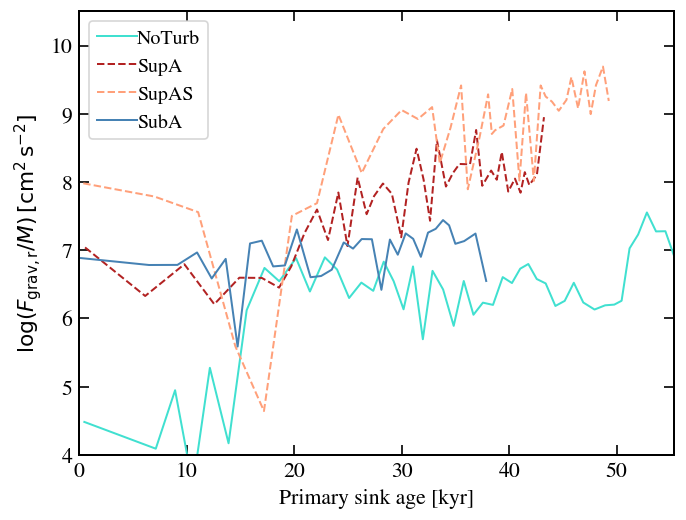}
    \caption{Evolution of angular momentum transported by the outflows (left panel), magnetic fields (middle panel) and by gravitation (right panel) within a cylinder of radius and height $1000$~AU.}
    \label{fig:fmaggrav}
\end{figure*}

We report three mechanisms for angular momentum transport to help accretion onto the central object. First, outflows are ubiquitous except in run \textsc{SupAS} where a monopolar and transient outflow develops (see Paper II). Second, spiral arms are present in the disk (similarly to \citealt{klassen_simulating_2016}) and exert gravitational torques that transport angular momentum outward. Third, magnetic braking occurs. We compute contributions to the angular momentum flux as  \citep{joos_protostellar_2012}
\be
F_\mathrm{out}=\left\lvert \int_\mathrm{S} \rho r v_\phi    \mathbf{v}  \cdot d\mathbf{S} \right\rvert
\ee
for the outflows (using the selection criteria presented in Paper II),
\be
F_\mathrm{grav}^\mathrm{r} =\left\lvert \int_\mathrm{S_{rad}} r \frac{ g_\phi }{4 \pi \mathrm{G}}   \mathbf{g}  \cdot d\mathbf{S} \right\rvert
\ee
for the gravitational torque and
\be
F_\mathrm{mag} = \left\lvert \int_\mathrm{S} r \frac{ B_\phi }{4 \pi}   \mathbf{B} \cdot d\mathbf{S}\right\rvert
\ee
for the magnetic torque. The first and third integrals are performed over the surface $\mathrm{S}$ of a cylinder centered onto the primary sink, of radius $R=1000$~AU and height $H=1000$~AU and is oriented along the angular momentum vector. The surface $\mathrm{S_{rad}}$ only accounts for the surfaces of this cylinder that are perpendicular to the cylindrical radial vector, since we expect gravitational torques to transport angular momentum radially rather than vertically. In order to have comparable values from one run and one timestep to the other, we divide those fluxes by the mass within the cylinder. Figure~\ref{fig:fmaggrav} shows $F_\mathrm{out}/M$, $F_\mathrm{mag}/M$ and $F_\mathrm{grav}^\mathrm{r}/M$ as a function of the sink age. Angular momentum transported by outflows is slightly larger in run \textsc{NoTurb} than in \textsc{SupA} and \textsc{SubA}. We find larger angular momentum transport from magnetic torques than from outflows. Magnetic braking is initially stronger in the initially most magnetized model (run \textsc{SubA}). After a sink age ${\sim}20$~kyr, it is smaller in run \textsc{SubA} than in \textsc{NoTurb}, suggesting that the turbulence reduces magnetic braking. This is confirmed by the even lower magnetic braking in runs \textsc{SupA} and \textsc{SupAS}. On the right panel of Fig.~\ref{fig:fmaggrav}, we observe that the gravitational torque is stronger with increasing turbulence. Overall, the magnetic torque generalles dominates the angular momentum transport, except in run \textsc{SupAS} at later times where magnetic and gravitational torques are comparable. We perform the same comparison with $R=100$~AU and the results remain qualitatively unchanged.

\subsection{Sink mass history and stellar multiplicity}
\label{sec:sinkmass}

\begin{figure*}
\centering
    \includegraphics[width=8cm]{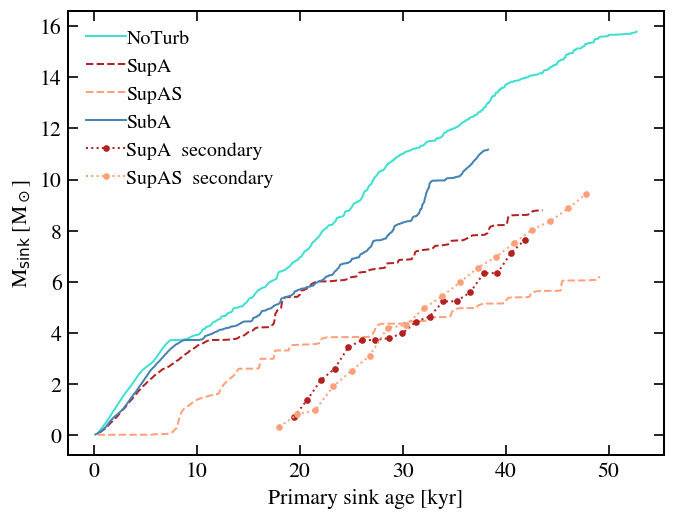}
    \includegraphics[width=8cm]{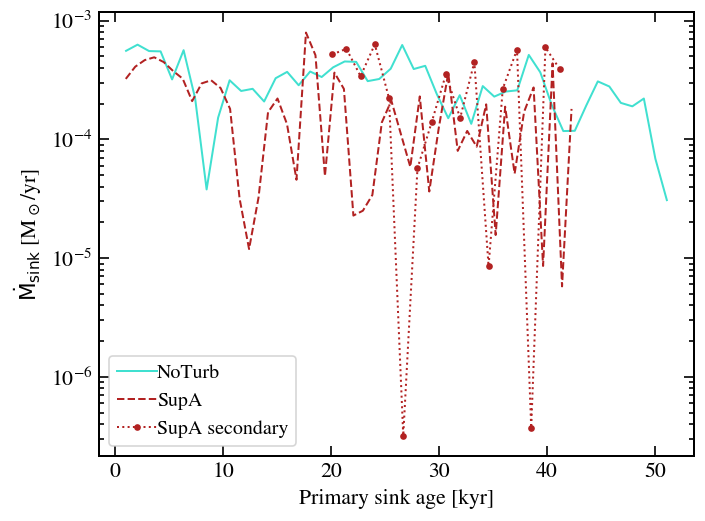}
    \caption{Primary and secondary sink mass (left panel) and accretion rate (right) as a function of primary sink age, for the four runs. The accretion rate is plotted for one subalfv\'enic (\textsc{NoTurb}) and one superalfv\'enic (\textsc{SupA}) run for readability. It has been smoothed over periods of ${\sim}1$~kyr.}
    \label{fig:t_msink}
\end{figure*}

\begin{figure*}
\centering
    \includegraphics[width=8cm]{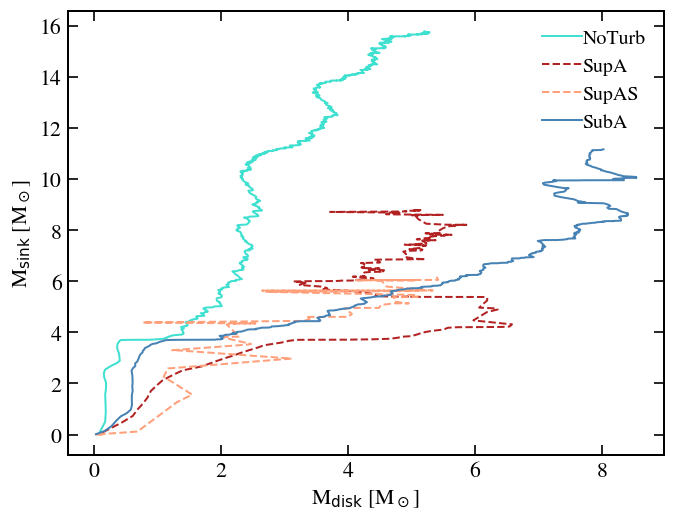}
    \includegraphics[width=8cm]{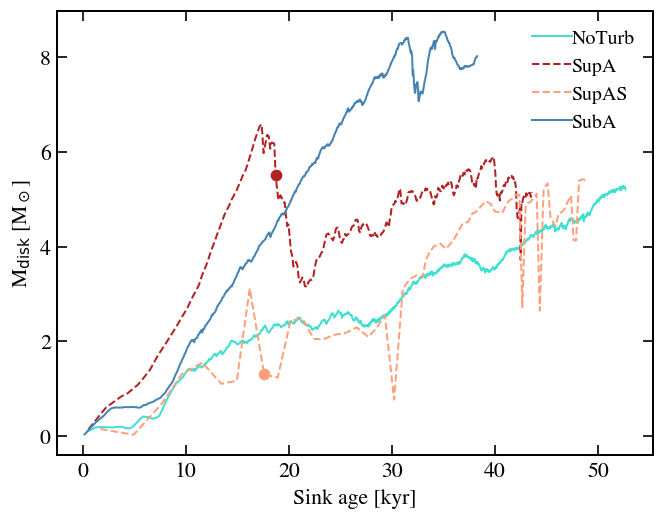}
    \caption{Primary sink mass against the disk mass (left panel) and disk mass as a function of time (right), for the four runs. In the right panel, coloured circles indicate the secondary sink formation epoch.}
    \label{fig:t_msink2}
\end{figure*}

To begin with, we study the mass evolution of the sink particles.
Left panel of Fig.~\ref{fig:t_msink} displays the primary and secondary (when there is) sink mass as a function of primary sink age.
The most massive star formed is $M\,{\simeq}\,15.8 \Msol$, in run \textsc{NoTurb}, at the end of the run when the primary sink age is $55$~kyr.
Globally, two different behaviours are visible, between subalfv\'enic (\textsc{NoTurb}, \textsc{SubA}) and superalfv\'enic runs (\textsc{SupA}, \textsc{SupAS}).
There is a delay of ${\sim}8$~kyr between runs \textsc{SupA} and \textsc{SupAS} but a comparable slope (mean accretion rate).
The mass accretion is much smoother in the subalfv\'enic cases than in the superalfv\'enic cases.
This is confirmed when looking at the accretion rate, displayed in the right panel of Fig.~\ref{fig:t_msink} for runs \textsc{NoTurb} and \textsc{SupA}.
It is smoothed over a temporal period of ${\sim}1$~kyr. 
The values for runs \textsc{SubA} and \textsc{SupAS} are not displayed here, for readability, and show similar features to runs \textsc{NoTurb} and \textsc{SupA}, respectively.
It is mainly between $10^{-4}$ and $10^{-3} \, \Msolyr$ in run \textsc{NoTurb}.
The accretion rate over the primary sink in run \textsc{SupA}, which includes initial turbulence, is first comparable to \textsc{NoTurb}.
After ${\sim} 12$~kyr it becomes erratic. Instantaneously (\textit{i.e.}, without the smoothing), it has most of the time zero values.
We recall that our sink accretion scheme relies on the presence of high-enough density and Jeans-unstable gas within the sink volume.
Hence, the absence of accretion, at a given time, means that the sink volume has not gathered enough material to be accreted.
The main accretion events in the superalfv\'enic runs are more dramatic than in the subalfv\'enic runs, with companion sink particles or orbiting massive clumps raising the primary sink mass by a fraction of a solar mass instantaneously. 
Averaged over time, we obtain accretion rates in agreement with observational values (\citealp{motte_high-mass_2018} and references therein).


We report the formation of a long-lived binary system in the two superalfv\'enic runs (\textsc{SupA} and \textsc{SupAS}, see Table~\ref{table:results}).
In run \textsc{SupAS}, three additional sink particles form from initial fragmentation and four in the disk plane, but merge with the primary or secondary sinks.
The secondary sink forms ${\approx} \,17$~kyr after the primary.
It occurs at the extremity of a spiral arm.
The secondary particle survives until the end of the run, \textit{i.e.} a lifetime ${\gtrsim}\,33$~kyr. 
This system also forms in the corresponding lower-resolution run (LR\textsc{SupAS}), at an age difference of ${\approx}18$~kyr instead of ${\approx}17$~kyr.
Thanks to a lower resolution, run LR\textsc{SupAS} has been carried out up to $t{\sim}106$~kyr and the secondary sink particle is ${\approx}60$~kyr old.
The stellar masses are then $9.8\Msol$ and $8.9\Msol$. 
The same formation mechanism leads to the birth of a long-lived companion in run \textsc{SupA}, at a primary sink age of ${\approx} \,19$~kyr.
Similarly, a binary system forms in LR\textsc{SupA} too but at later times (${\approx} \,43$~kyr after the primary), with final masses of $19.7\Msol$ and $8\Msol$ at $t\, {\approx}\, 102$~kyr. 
The secondary sink is $30$ kyr old.
Interestingly, after $t\, {\sim}\,78$~kyr, the primary sink gains only a fraction of a solar mass within more than $20$~kyr, while the companion accretes $6\Msol$.
Overall, we obtain long-lived (at least tens of kyr) binary systems in the superalfv\'enic runs.
This demonstrates the effect of turbulence on fragmentation, even when magnetic fields are present at a moderate level.

The sinks forming the binary system in runs \textsc{SupA} and \textsc{SupAS} have mass ratios of the order of unity (in run LR\textsc{SupA} where it may tend towards $1$), and even greater than $1$ in run \textsc{SupAS}.
It can be seen in the sink mass history (left panel of Fig.~\ref{fig:t_msink}) that the primary sink is partially starved due to the presence of the secondary sink, as compared to the subalfv\'enic runs where no binary forms.
In both runs, the secondary sink mass quickly becomes comparable to (and even greater than, in run \textsc{SupAS}) the primary sink mass.
The evolution of the secondary sink mass is very similar in runs \textsc{SupA} and \textsc{SupAS}, and the accretion rate is greater than for the primary sink (right panel of Fig.~\ref{fig:t_msink}).
This suggests that the primary sink accretion disk could be a more favorable place for accretion than the central location of the primary sink (when it dominates the total sink mass). 
Indeed, the radial gravito-centrifugal equilibrium in the disk likely reduces the accretion rate onto the primary sink but not onto the secondary sink. 

As mentioned above, the binary systems form from disk fragmentation (see Sect.~\ref{sec:diskfrag}) rather than core fragmentation.
All the sink particles formed from initial core fragmentation have merged.
We recall that we use AMR based on the thermal Jeans length. 
The numerical convergence we perform with LR runs advocates for a physical fragmentation, rather than numerical. 
Nevertheless, the absence of a criterion to prevent sink merging in our simulations means that the final sink number may be higher.
Hence, our multiplicity results can be considered, to first order, as lower-limit values.

Fig.~\ref{fig:t_msink2} shows the relation between the primary sink and the disk mass (left panel; the proper disk mass evolution is displayed in the right panel and discussed in Sect.~\ref{sec:diskmass}).
It appears that there is a correlation between several sink mass gain events and disk mass loss events in the four runs, but the masses involved are much smaller in the subalfv\'enic runs.
This is consistent with the gas falling smoothly onto the central star via the accretion disk, in subalfv\'enic runs, while in superalfv\'enic runs, clumps form in the disk and are subsequently accreted.


\subsection{Primary disk properties}
\label{sec:diskppt}

\subsubsection{Mass and radius}
\label{sec:diskmass}

The disk mass temporal evolution is displayed in the right panel of Fig.~\ref{fig:t_msink2}. 
It globally increases with sink age, with more variations in the superalfv\'enic runs. 
We obtain disk masses ranging from ${\approx}1-8\Msol$ (for $t>10$~kyr). 
This confirms the trend observed in C21 and extends it to a turbulent medium: in the hydrodynamical case, disks are more massive ($10\Msol$) than in the presence of magnetic fields. 


Figure~\ref{fig:t_mrdisk} shows the primary disk radius as a function of time (left panel) and its comparison with the analytical prediction of \cite{hennebelle_magnetically_2016}.
As shown in the left panel of Fig.~\ref{fig:t_mrdisk}, the disk radius is most of the time between $50$ and $200$~AU in all runs.
Large and ponctual increases coincide with the presence of a large spiral arm.
The quasi-periodic variations found in runs \textsc{SupA} and \textsc{SupAS}  are due to the orbital motions which impact the gas dynamics between the two stars.

\subsubsection{Semi-analytical estimate of the disk radius}
\label{sec:disksize}

\begin{figure*}
\centering
    \includegraphics[width=8cm]{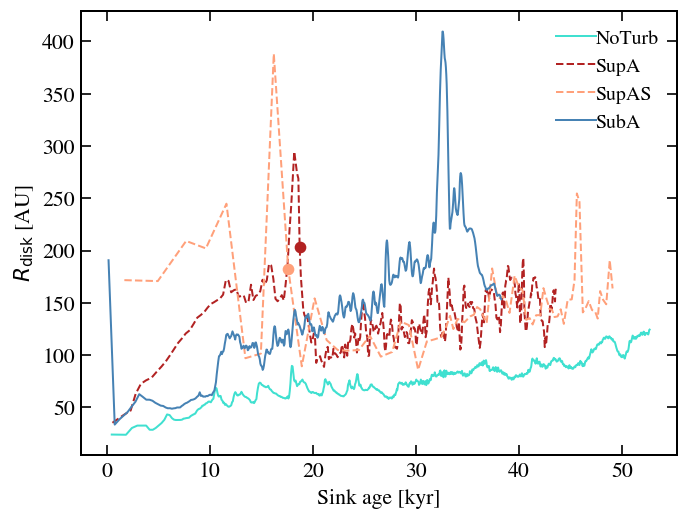}
    \includegraphics[width=8cm]{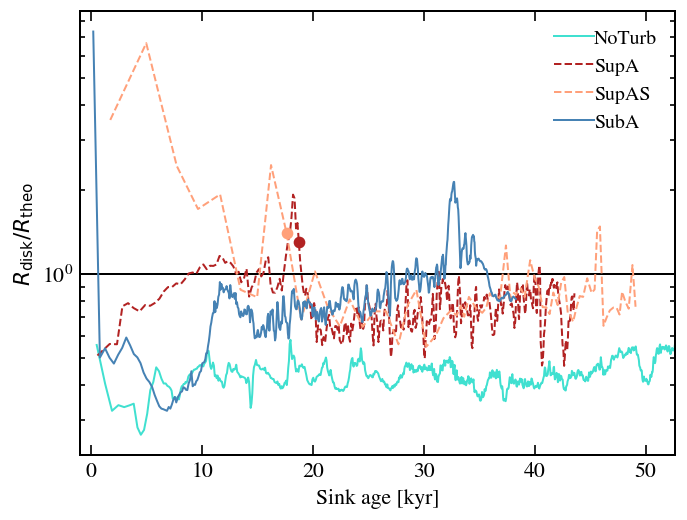}
    \caption{Disk radius (left panel) and ratio between the disk radius and the theoretical value (right, Eq.~\ref{eq:rd_ad}) as a function of time, for the four runs. Coloured circles indicate the secondary sink formation epoch.}
    \label{fig:t_mrdisk}
\end{figure*}

We compare the disk sizes with the theoretical predictions from \cite{hennebelle_magnetically_2016} for magnetically-regulated disk formation with ambipolar diffusion.
Those are obtained from the equality between various timescales at the centrifugal radius.
On the one hand, the timescale to generate toroidal field from differential rotation and the timescale for ambipolar diffusion to diffuse it vertically.
On the other hand, the magnetic braking and the rotation timescales are set equal as well.
The disk radius set by ambipolar diffusion is then
\be
r_\mathrm{d,AD} {\simeq} 18 \mathrm{AU} \times \delta^{2/9} \left(\frac{\eta_\mathrm{AD}}{0.1 \mathrm{s}} \right)^{2/9} 
\left(\frac{B_z}{0.1 \mathrm{G}} \right)^{-4/9}
\left(\frac{M_\mathrm{d}+M_\mathrm{\star}}{0.1 \Msol}\right)^{1/3},
\label{eq:rd_ad}
\ee
where $\delta$ is the ratio between the initial density profile and the singular isothermal sphere (SIS, \citealt{shu_self-similar_1977}), and $M_\mathrm{d}$ is the disk mass.
By comparing our density profile to the SIS, we take $\delta=10$, in agreement with \cite{hennebelle_collapse_2011}, and the mean magnetic field strength within the disk as a proxy for the component $B_z$.
For $\eta_\mathrm{AD}$ we take the azimuthally-averaged value at the cell located further away from the sink, within the disk selection.

The disk sizes agree within a factor of ${\approx}2-3$ with the prediction above (right panel of Fig.~\ref{fig:t_mrdisk}).
We find roughly similar disk radii in all runs (Fig.~\ref{fig:t_mrdisk}), in agreement with this model predicting that the disk radius does not explicitly depend on the amount of angular momentum available (in opposite to the hydro case).
Hence, the disk around primary sinks appears to be set by magnetic regulation.


The deviation from the analytical prediction is similar to what has been found in C21: the disk radius is analytically slightly overestimated for $\mu=2$ and $\mu=5$ runs, and underestimated when there is a significant rotational support (our \textsc{SupA} and \textsc{SupAS} runs, and the run MU5ADf in C21 with $\mu=5$ and $E_\mathrm{rot}/E_\mathrm{grav}=5\%$).
This underestimation may be due to the presence of spiral patterns in disks when there is strong rotation/turbulent support, as the disk becomes gravitationally unstable.

\subsubsection{Column density maps}

\begin{figure*}
\centering
    \includegraphics[width=16cm]{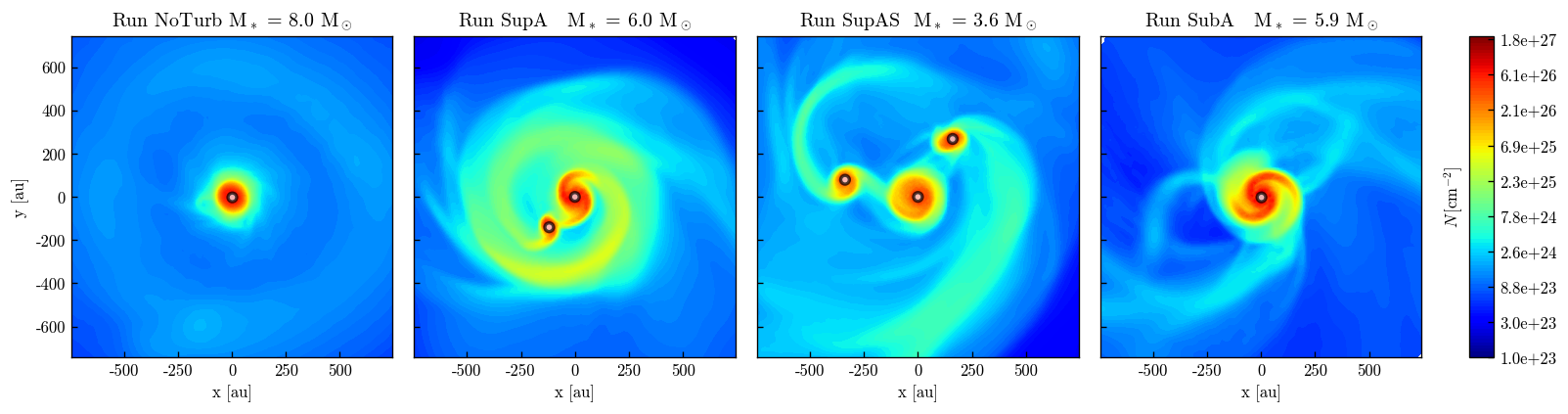}
    \caption{Column density maps at $t{\approx}50$~kyr. The integration is done along the angular momentum vector direction. From left to right: run \textsc{NoTurb}, \textsc{SupA}, \textsc{SupAS}, \textsc{SubA}. White dots represent sink particles.}
    \label{fig:coldens_paperII}
\end{figure*}

\begin{figure}
\centering
    \includegraphics[width=8cm]{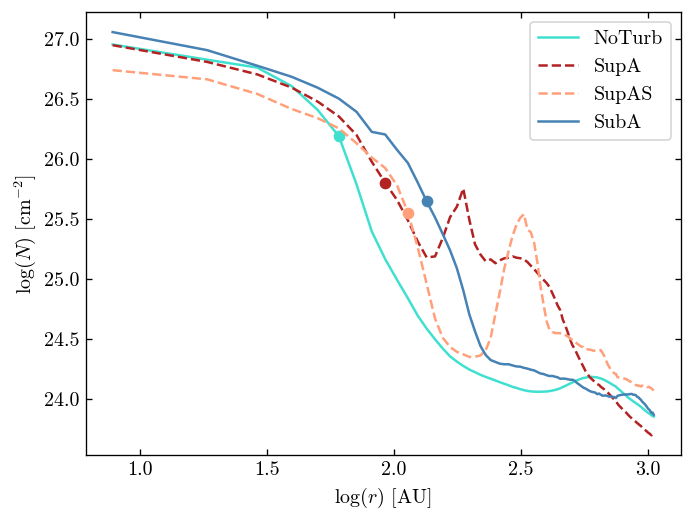}
    \caption{Azimuthal median of the column density as a function of the radius, for the four runs, at $t=50$~kyr. Coloured circles indicate the radius derived from our disk definition.}
    \label{fig:r_nmedian}
\end{figure}

Our disk definition is physically motivated (see Sect.~\ref{sec:diskdef}) and allows for a comparison with previous numerical studies.
Nevertheless, even though Keplerian disks can be traced from the position-velocity diagram (\textit{e.g.} \citealt{cesaroni_study_2005}), the other criteria can be hardly checked observationally. For that purpose, Fig.~\ref{fig:coldens_paperII} gives the column density maps at $t=50$~kyr. The individual disks have a column density larger than their surroundings by at least one order of magnitude. Spiral arms are visible, as well as the presence of circumbinary gas. Figure~\ref{fig:r_nmedian} shows the azimuthal median of the column density as a function of the radius.
Coloured circles indicate the radius which has been determined using our disk definition. We choose a median rather than a mean in order to reduce the impact of non-axisymetries due to fragments or sinks. A smooth but visible transition (in slope and values) is observed around the disk radius we have derived. For runs \textsc{SupA} and \textsc{SupAS}, the dense circumbinary gas makes more difficult the primary disk determination.

\subsubsection{Disk fragmentation}
\label{sec:diskfrag}

As shown above, the origin of sink formation is disk fragmentation.
This occurs in high-density gas at the extremity of spiral arms, similarly to the non-magnetic case studied in \cite{mignon-risse_new_2020}.
Spiral arms are found in all runs, and at (nearly) all times.
We check for the Toomre parameter (see \citealt{kratter_gravitational_2016} for a review on disk fragmentation) value before the development of spiral arms, as it indicates the unstable state to axisymmetric perturbations which lead to spiral arm development (often done in the literature, see \textit{e.g.}, \citealt{kratter_fragmentation_2006}, \citealt{klassen_simulating_2016}, \citealt{ahmadi_core_2018}).
We compute it as in \cite{mignon-risse_new_2020} - we do not display it here for conciseness. 
Disks are Toomre-unstable with typically $Q\lesssim0.7$, except in their outer parts ($r\gtrsim100$~AU).
Let us note that, as shown in Figs.~\ref{fig:betamaps} and ~\ref{fig:vprofiles}, disks are (largely) dominated by thermal pressure rather than magnetic pressure ($\beta >1$). Hence, computing the thermal Toomre (without the magnetic pressure) is sufficient.

However, the presence of spiral arms does not indicate fragmentation neither the formation of a multiple stellar system.
In run \textsc{NoTurb}, no fragment forms. In run \textsc{SubA}, the first fragments form at a sink age around ${\sim}32$~kyr but fall and merge back onto the primary disk. In run \textsc{SupA} the first fragment formation occurs at a sink age ${\sim}18$~kyr and in run \textsc{SupAS} before $16$~kyr.
These differences suggest that the interplay of turbulence and magnetic fields may impact disk fragmentation and sink formation.

Several criteria have been frequently used in the litterature to address the origin of disk fragmentation, in addition to the Toomre $Q$ parameter.
\cite{klassen_simulating_2016} used the Gammie criterion \citep{gammie_nonlinear_2001}, aiming at comparing the cooling time to the orbital time. Even though we find that the local cooling time is generally smaller than the orbital time (even in our non-fragmenting run \textsc{NoTurb}, in agreement with \citealt{klassen_simulating_2016} and using the same procedure), radial radiative flux is propagating in the disk to heat it up at a similar rate. Therefore, cooling is not responsible for fragmentation here. 
Indeed, the Gammie model is well adapted to disks possessing local cooling processes while our disks heat and cool radiatively.
Moreover, \cite{gammie_nonlinear_2001} hypothesizes the disk to be cool and very thin, while we get an aspect ratio of typically $0.1-0.2$. 
Furthermore, his local model is only applicable if $c_\mathrm{s}/(r \Omega) \ll 0.12$ (Eq.~26 of \citealt{gammie_nonlinear_2001}) where $\Omega$ is the orbital frequency. With $c_\mathrm{s} {\approx}10^5 \, \mathrm{cm \, s^{-1}}$, $r {\approx}100 \, \mathrm{AU}$ and $\Omega {\sim}10^{-11} \, \mathrm{s^{-1}}$, we get $c_\mathrm{s}/(r \Omega) {\sim} 10 \gg 0.12$. Hence, we argue that the model of \cite{gammie_nonlinear_2001} does not apply to the massive protostellar disks formed in this study.

\begin{figure}
\centering
    \includegraphics[width=8cm]{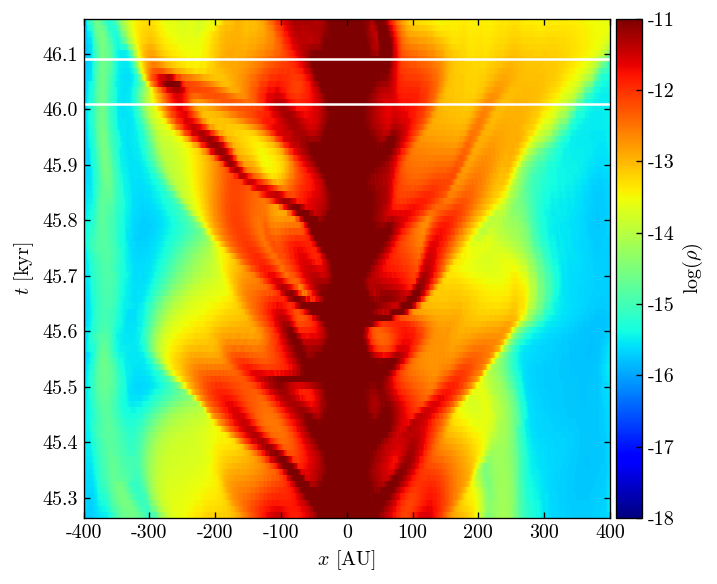}
    \caption{Modified logarithmic colormap of the density as a function of the position and time, along a given direction in the disk plane in run \textsc{SupA}. Spiral arms appear as diagonal lines of enhanced density. We are particularly interested in the spiral arm collision occurring at $t=46.05$~kyr, indicated by the two horizontal lines located at $\pm0.04$~kyr from the collision time.}
    \label{fig:t_spiral}
\end{figure}

\begin{figure}
\centering
    \includegraphics[width=8cm]{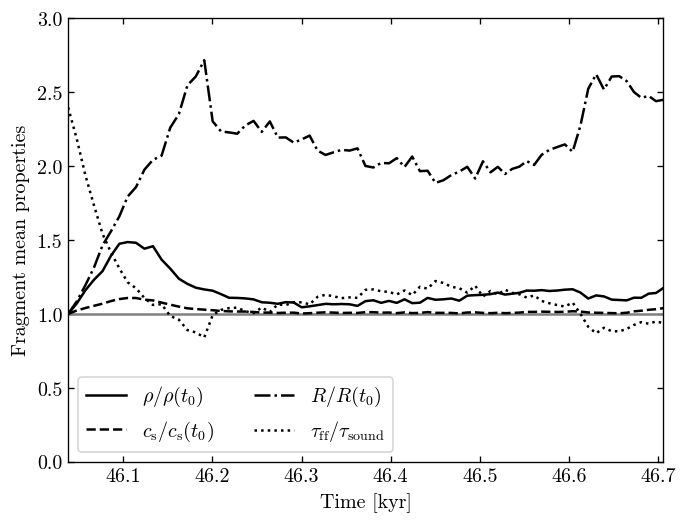}
    \caption{Properties of the fragment leading to secondary sink formation ($t=46.71$~kyr) in run \textsc{SupA}. The density, radius, and sound speed are normalized to the value at the formation time. All quantities are averaged over the fragment volume.}
    \label{fig:t_fragppties}
\end{figure}

We investigate spiral arm collision as a possibility to form fragments. Figure~\ref{fig:t_spiral} shows the gas density as a function of position (along a fixed direction in the disk plane) and time in run \textsc{SupA}. Spiral arms are visible at all times as diagonal lines of enhanced density in the $[x,t]$ plane. A spiral arm collision event is indicated by the two horizontal lines (occurring in a similar fashion as in \citealt{oliva_modeling_2020}). It creates a region of enhanced density, \textit{i.e.} a fragment, and decouples it from its parent arm. Collisions are observed in runs \textsc{SubA} and \textsc{SupAS} too. This process is favored when the spiral arms can sufficiently extend radially away from the primary star, which does not occur in run \textsc{NoTurb}. The rapid growth of the spiral arms in the turbulent runs can be seen in the left panel of Fig.~\ref{fig:t_mrdisk}. When turbulence is included, it brings additional angular momentum for spiral arms to extend (\citealt{hennebelle_spiral-driven_2016}, \citealt{hennebelle_spiral-driven_2017}). The initial distribution of angular momentum computed with respect to the center of the domain increases with the distance. Thus, the less turbulent the core is, the longer it takes for gas with a similar angular momentum to reach the center. This explains why fragments form earlier in run \textsc{SupAS} than in runs \textsc{SupA} and \textsc{SubA} (see below a comparison between runs \textsc{SupA} and \textsc{SubA}). To conclude, we find disk fragmentation is due to spiral arm collision and favored by turbulence.

In order to understand the collapse of a fragment, let us follow the properties of the fragment of \textsc{SupA} presented above. Cells with a density larger than $10^{-12} \mathrm{g\, cm^{-3}}$ and a distance to the primary sink larger than $250$~AU (to avoid the disk) are selected as part of the fragment. Figure~\ref{fig:t_fragppties} shows the mean density, mean sound speed, size and ratio of the free-fall time to the thermal sound-crossing time of the fragment, from its formation time ($t=46.05$~kyr) to sink formation time ($t=46.71$~kyr). The free-fall time is computed as in Eq.~\ref{eq:tauff} with the fragment mean density, and the sound-crossing time is equal to $2R/c_\mathrm{s}$ with $2R$ the estimate on the fragment diameter and $c_\mathrm{s}$ its mean sound speed.
This ratio gives an order-of-magnitude estimate of the fragment's stability to density perturbations: a ratio smaller than one indicates it is gravitationally unstable. This estimate only accounts for thermal pressure because it largely dominates magnetic and radiative pressures locally. 
While the sound speed and density are found roughly constant with time, the fragment radius increases by a factor ${\sim}2.5$ compared to its initial value. The fragment has accreted mass. Moreover, those quantities directly dictate the evolution of the ratio between the free-fall time and sound-crossing time. Indeed, it is proportional to $c_\mathrm{s}/(\sqrt{\rho} R)$, thus it tends towards an unstable state as $R$ increases. Figure~\ref{fig:t_fragppties} shows it is slightly unstable during two distinct epochs. The sink forms during the second epoch, indicating that part of the fragment has become bound, in addition to being Jeans unstable.

The previous analysis on fragment formation does not allow us to distinguish runs \textsc{SubA} and \textsc{SupA}, \textit{i.e.} if there is an impact from magnetic fields on disk fragmentation. The fragments formed in \textsc{SubA} merge back onto the disk and \textsc{SupA} produces a companion, while both runs have a same level of turbulence and therefore, of angular momentum. Moreover, the first fragment forms significantly later in run \textsc{SubA} than in \textsc{SupA}. These two observations can be explained by the magnetic braking being stronger in \textsc{SubA} than in \textsc{SupA} (see Fig.\ref{fig:fmaggrav} and Sect.~\ref{sec:alignangmom}). It takes longer for larger-scale gas - which carries more angular momentum - to fall onto the disk and spiral arms, delaying the first fragment formation. Similarly, magnetic braking slows down the gas rotation and makes the fragment formed in \textsc{SubA} fall onto the disk.

To sum up, turbulence brings additional angular momentum to the ubiquitous spiral arms, which do not grow significantly in the non-turbulent run. This additional angular momentum favors their radial extension, their subsequent collision with another spiral arm and therefore the formation of fragments. When turbulence is subalfvenic, magnetic braking delays this process and drives the inward migration of fragments, preventing long-lived stellar systems to emerge.

\subsubsection{Characteristic velocities and magnetic field components}
\label{sec:diskvel}

\begin{figure*}
\centering
    \includegraphics[width=7cm]{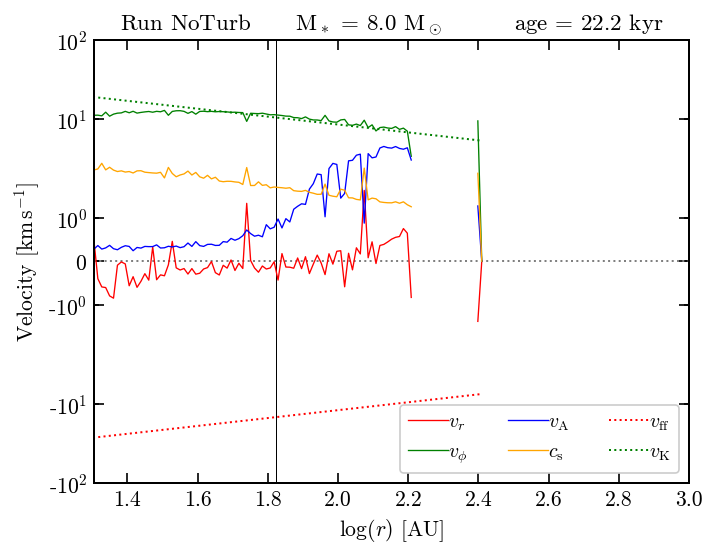}
    \includegraphics[width=7cm]{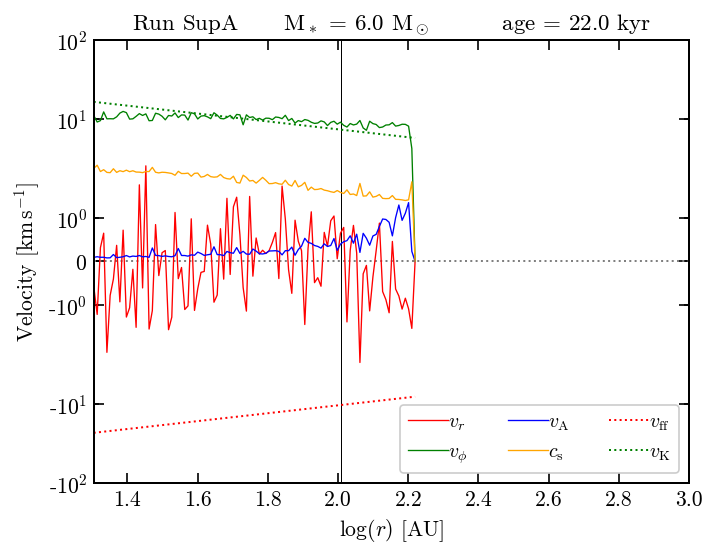}
    \includegraphics[width=7cm]{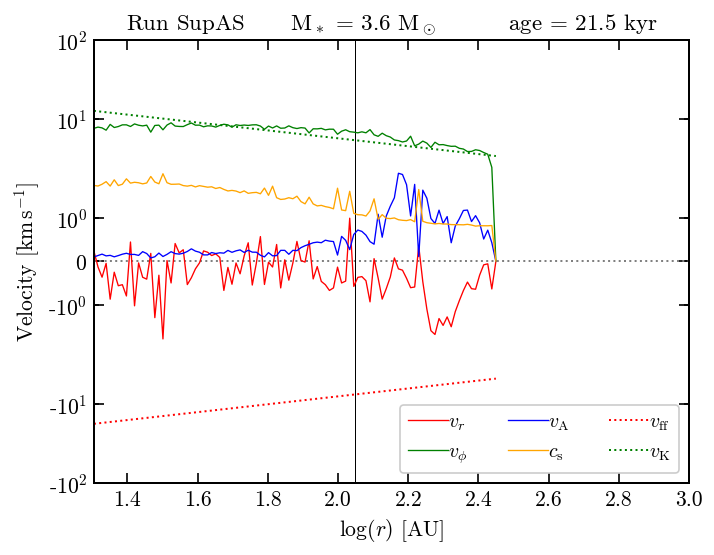}
    \includegraphics[width=7cm]{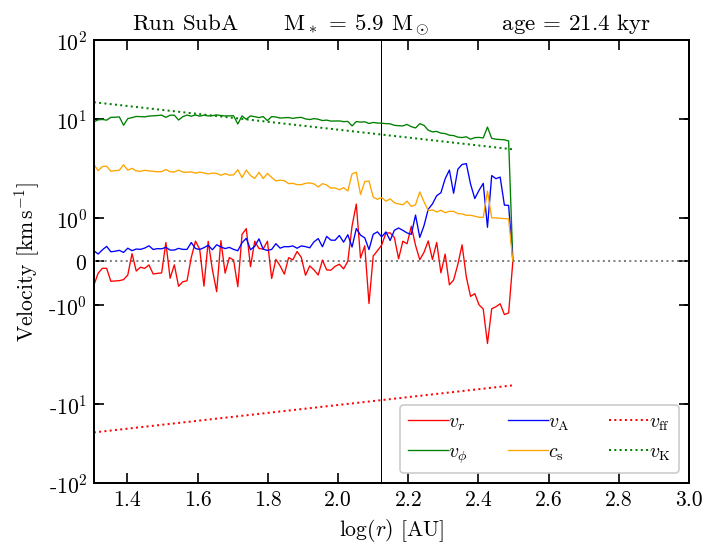}
    \caption{Azimuthally-averaged radial and azimuthal velocities, Alfv\'en speed, isothermal sound speed, free-fall velocity and Keplerian velocity as a function of the radius at $t=50$~kyr in the disk selection. The vertical line indicates the disk radius plotted in Fig.~\ref{fig:t_mrdisk} (which encapsulates $90\%$ of the disk mass). Top row: run \textsc{NoTurb} (left), \textsc{SupA} (right). Bottom row: run \textsc{SupAS} (left), \textsc{SubA} (right). Discontinuities are due to the disk selection being on a cell-by-cell basis, without connectivity criterion.}
    \label{fig:vprofiles}
\end{figure*}

\begin{figure*}
\centering
    \includegraphics[width=7cm]{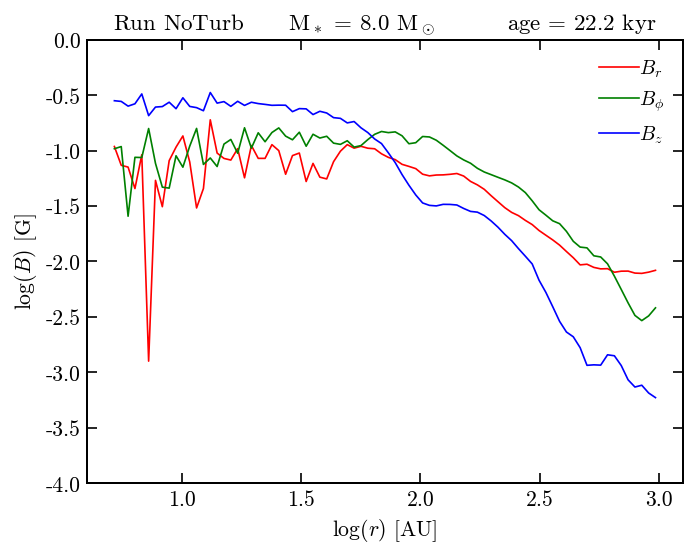}
    \includegraphics[width=7cm]{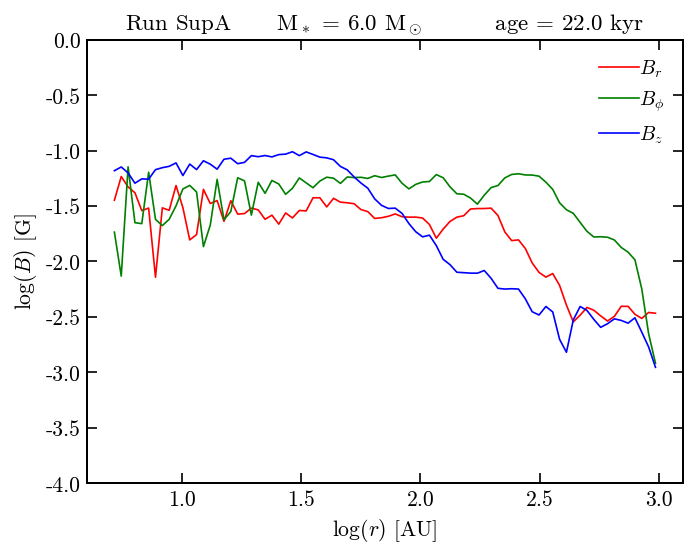}
    \includegraphics[width=7cm]{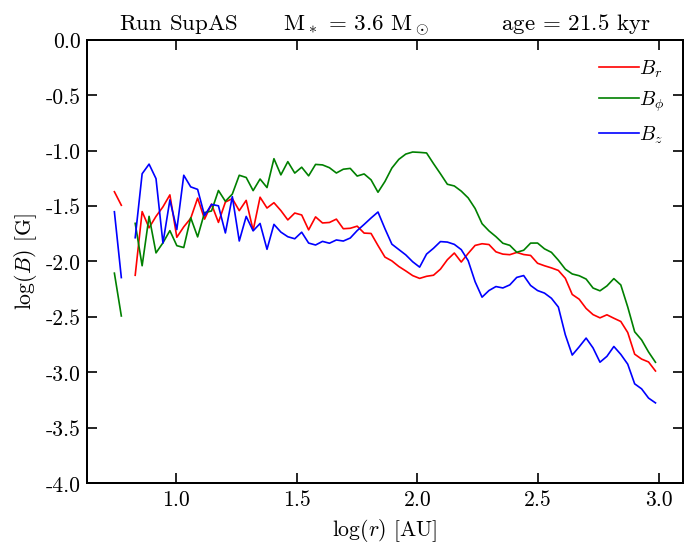}
    \includegraphics[width=7cm]{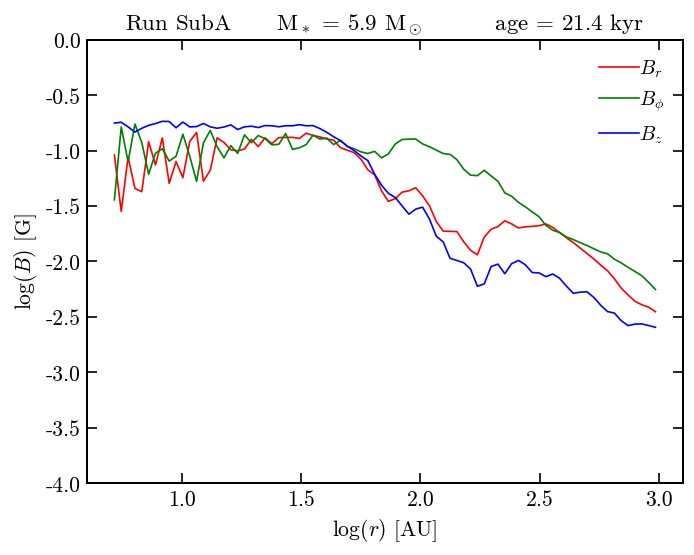}
    \caption{Azimuthally-averaged magnetic field components as a function of the radius at $t=50$~kyr. Top row: run \textsc{NoTurb} (left), \textsc{SupA} (right). Bottom row: run \textsc{SupAS} (left), \textsc{SubA} (right).}
    \label{fig:Bprofiles}
\end{figure*}

Figure~\ref{fig:vprofiles} shows the radial profile of the azimuthally-averaged characteristic velocities in the disk selection, using the disk plane as the $(\mathbf{u}_r,\mathbf{u}_\theta)$ plane.
Overall, the azimuthal velocity is in agreement with a Keplerian profile.
It is slightly super-Keplerian in turbulent runs (\textsc{SupA}, \textsc{SupAS}, \textsc{SubA}) at radii $\gtrsim 60$~AU.
In all runs, it becomes sub-Keplerian as the radius decreases.
This is due to the gravitational field being dominated by the central object and diminished by the sink softening length mentioned in Sect.~\ref{sec:resol}.
The disks are roughly Keplerian, hence the infall velocity is much smaller than the free-fall velocity and typically smaller than $1\mathrm{\, km \, s^{-1}}$.
The rotation motions, and infall motions beyond the disk, are supersonic.
The cells in the primary disk plane in the binary systems appear close to Keplerianity up to ${\sim}1000$~AU (not shown here), when measured at $t=50$~kyr (\textit{i.e.} when the secondary sink is much less massive than the primary sink).
In the absence of a strong stellar activity from one component, they could be identified as large disks \citep{johnston_keplerian-like_2015}.

As shown in the last row of Fig.~\ref{fig:betamaps}, all primary disks have plasma beta $\beta>1$(thermally-dominated).
From density maps and Fig.~\ref{fig:vprofiles} we observe that the disk radius is close to the point where a change from thermally-dominated to magnetically-dominated region is observed.
We argue that a physically-motivated criterion for the identification of individual disks is the plasma beta with $\beta \gtrsim 1$ (equivalent to $P_\mathrm{th}\gtrsim P_\mathrm{mag}$), since $c_\mathrm{s}/v_\mathrm{A} = \sqrt{\gamma \beta/2} {\sim} \sqrt{\beta}$, especially at late times. 
In our simulations, it encapsulates that the disk size is regulated by ambipolar diffusion, in contrast to the ideal MHD case (\citealt{masson_ambipolar_2016}, C21).
This criterion only is not sufficient though because of the existence of thermally-dominated ($\beta>1$) filaments, as well as parts of the circumbinary disk (see Sect.~\ref{sec:circumd}), but it works well in the vicinity of the stellar object.
The filaments can be discarded by an additional criterion based on rotation.

Figure~\ref{fig:Bprofiles} displays the azimuthally-averaged magnetic field components using the same coordinates as in Fig.~\ref{fig:vprofiles}.
We select cells in the disk plane but these are not restricted to the disk selection in order to probe the outer regions too.
Strikingly, the vertical component $B_z$ dominates for $r\lesssim 50$~AU in runs \textsc{NoTurb} and \textsc{SubA}, and in run \textsc{SupA} most of the computational time.
A dominantly poloidal magnetic field is a necessary condition for launching centrifugal jets \citep{blandford_hydromagnetic_1982}. 
In runs \textsc{SupA} and \textsc{SupAS}, the magnetic field components show more variations with respect to time than in runs \textsc{NoTurb} and \textsc{SubA}. We observe many occurrences of $B_z$ and $B_r$ having opposite evolutions.
Changes between $B_z$ and $B_r$ could be explained by the orbital motion of the primary sink, which occurs at  superalfvenic velocities. A change from $B_z$ to $B_r$ can be attributed to magnetic field lines lagging behind the sink, while ambipolar diffusion would favor a return to $B_z$, as in runs \textsc{NoTurb} and \textsc{SubA}.
Following the method from C21, we define the orbital Elsasser number for ambipolar diffusion $\mathrm{Am}$ as the ratio between the orbital time $\tau_\mathrm{orb}$ and the ion-neutral collision time $t_\mathrm{in}=\eta_\mathrm{AD} \rmc^2/(4\pi v_\mathrm{A}^2)$
\be
\mathrm{Am} = \frac{4 \pi v_\mathrm{A}^2  \tau_\mathrm{orb}}{\rmc^2 \eta_\mathrm{AD}}.
\ee
Taking the values in the vicinity of the primary sink we have $v_\mathrm{A} {\sim}0.1 \mathrm{km \, s^{-1}}$ (see Fig.~\ref{fig:vprofiles}), $\eta_\mathrm{AD}=0.1$, and $\tau_\mathrm{orb} {\simeq}1$~kyr (Fig.~\ref{fig:t_mrdisk}), which gives $\mathrm{Am} {\sim} 0.4$ which is of the order of unity (recall that it is highly-dependent on $v_\mathrm{A}$, which increases away from the sink) and indicates that, indeed, kinematical effects compete with ambipolar diffusion.

At larger radii, including within the disk radius, the toroidal component $B_\phi$ dominates in all runs.
This is due to the magnetic field lines being twisted by the disk rotation. 
Eventually, the radial component $B_r$ dominates at even larger radii.
It has been produced by the magnetized, collapsing gas (and the streamers), pulling the field lines which fan out at infinity and form an hour-glass shape (\textit{e.g.}, \citealt{galli_collapse_1993}).

\subsection{Secondary disk (runs \textsc{SupA}, \textsc{SupAS})}
\label{sec:seconddisk}

\begin{figure*}
\centering
    \includegraphics[width=7cm]{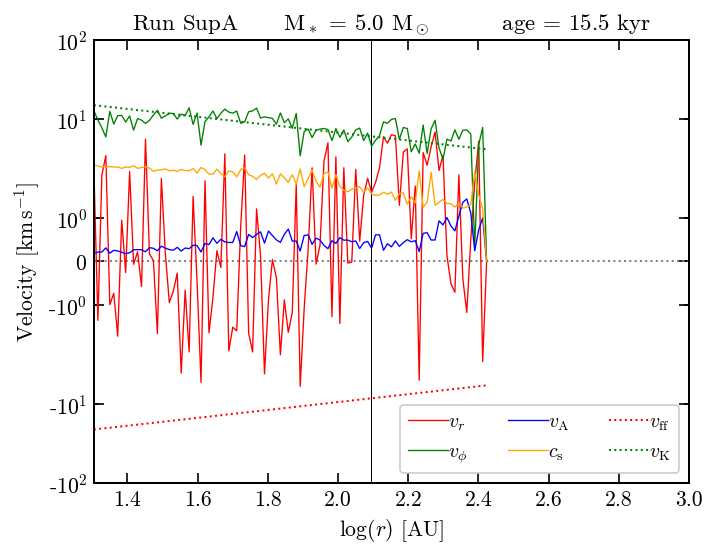}
    \includegraphics[width=7cm]{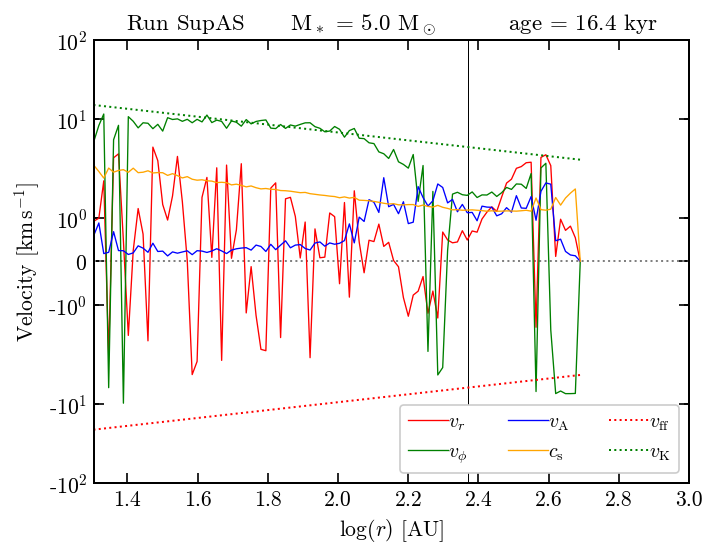}
    \caption{Same as Fig.~\ref{fig:vprofiles_2} for the secondary sinks in runs \textsc{SupA} (left) and \textsc{SupAS} (right) when their mass is $5\Msol$.}
    \label{fig:vprofiles_2}
\end{figure*}

Let us focus on the disk around secondary sinks in run \textsc{SupA} and \textsc{SupAS}, in comparison with the primary disk in the same run (Sect.~\ref{sec:diskppt}).
First, they rotate in the same direction as the primary disks.
Figure~\ref{fig:vprofiles_2} shows the radial profile of the azimuthally-averaged characteristic velocities within the cells corresponding to our disk selection criteria applied to the secondary sink environment.
Similarly to the primary disks, their rotation profile is consistent with Keplerian rotation and they have $c_\mathrm{s}>v_\mathrm{A}$.
Their plasma beta is shown in Fig.~\ref{fig:circumbeta}, (which is a slice centered on the center of mass of the two sinks) and shows how similar the two disks are.
The region where there is an inversion in the azimuthal velocity $v_\phi$ corresponds to the closest part of the primary disk which dominates the azimuthal average (seen also around the primary sinks at later times than displayed in Fig.~\ref{fig:vprofiles}). This also gives an upper-limit to the secondary disk radius, in complement to the transition radius at which $\beta {\sim}1$.

In both runs, the toroidal component of the magnetic field dominates most of the secondary disk region.
At small radii ($<50$~AU), the dominant component is not constant with time, as seen around the primary sinks (Sec.~\ref{sec:diskvel}).
The co-evolution of $B_z$ and $B_r$ likely results from the same mechanism, \textit{i.e.} a $B_z$ component increased by ambipolar diffusion but turning into $B_r$ as the secondary sink moves at a superalfvenic speed.
Nonetheless, the temporal evolution does not show a clear pattern permitting us to link the aforementionned observations with characteristic timescales (\textit{e.g.} the orbital period).
By the end of the run \textsc{SupA}, the secondary disk becomes dominated by $B_z$ most of the time, similarly to the primary disks in runs \textsc{NoTurb}, \textsc{SupA} and \textsc{SubA}. While the long-term evolution of this system is beyond the scope of this study, one may expect a magnetic structure favorable to outflow launching to build more rapidly around the secondary sink in run \textsc{SupA} than in run \textsc{SupAS}.

As can be seen on density maps (not shown here) and velocity profiles (\textit{e.g.} taking the radius at which $v_\phi$ drops as an upper-limit, in Fig.~\ref{fig:vprofiles_2}), the secondary disk radius is found to be of the order of ${\sim}100-150$~AU, in agreement with the transition radius $\beta {\sim}1$.
This is the same order of magnitude as found for the primary disks, and consistent within a factor of $2$ with magnetic regulation (Eq.~\ref{eq:rd_ad} predicts ${\sim}200$~AU in these cases).

\subsection{Circumbinary disk (runs \textsc{SupA}, \textsc{SupAS})}
\label{sec:circumd}

\begin{figure*}
\centering
    \includegraphics[width=8cm]{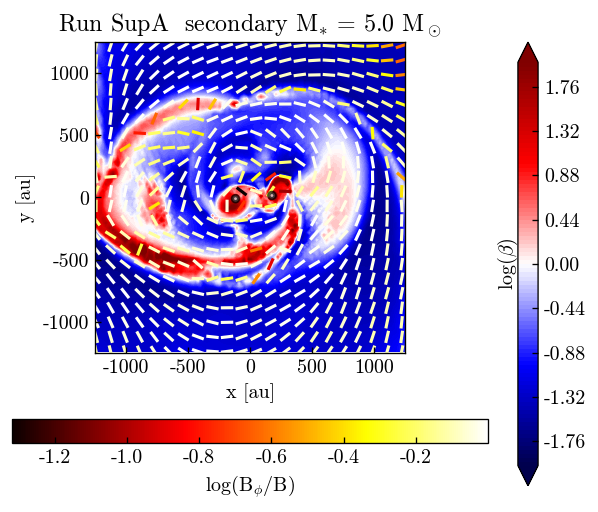}
    \includegraphics[width=8cm]{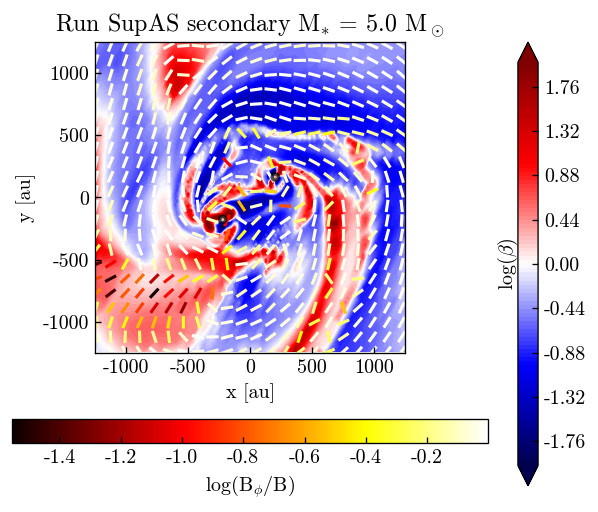}
    \caption{Plasma beta slices of $2500$~AU in the rotation plane, centered onto the center of mass, when the secondary sink mass is $5\Msol$. The magnetic field topology is overplotted, white colors corresponding to a dominantly toroidal field. Left panel: run \textsc{SupA}; right panel: run \textsc{SupAS}.}
    \label{fig:circumbeta}
\end{figure*}

Runs \textsc{SupA} and \textsc{SupAS} show the formation of binary systems.
The primary disks are embedded in a rapidly evolving circumbinary disk-like structure.

We investigate whether the binary separation is consistent with hydrodynamical disk radii, whose size is set by the centrifugal barrier from initial angular momentum conservation, rather than magnetic regulation.
Nonetheless, this calculation will lead to a constant value, while we observe elliptic orbits (with a factor of ${\sim}2$ between the periastron radius and the apastron radius) whose distance increases with time, once integrated over several orbits. 
The core rotational energy is $E_\mathrm{rot} = \mc ( \rc \Omega)^2/2 = \mc (J/\mc)^2 / (2 \rc^2)$, where $\rc$ is the its radius and $J=\norm{ \int_{r<\rc} \mathbf{r} \times \rho \mathbf{v} \, \mathrm{d}V}$ its angular momentum, centered onto the primary sink.
Equalling the rotational energy and the gravitational energy $3\rmg \mc/(5R)$ (assuming a uniform density) we obtain the hydro disk radius
\be
r_\mathrm{d,hy}\, {\simeq} \, 100 \mathrm{AU} \times 
\left( \frac{J/\mc}{5 \times 10^{21} \mathrm{\, cm^2 \, s^{-1}}} \right)^2 \left( \frac{\mc}{100 \Msol} \right)^{-1}.
\ee
The quantity $J/\mc$ is displayed in Fig.~\ref{fig:dx_j} as a function of the integration radius, here we take the value for $R=\rc$.
For run \textsc{SupA}, $J/\mc \,{\simeq} \, 8 \times 10^{21} \mathrm{\, cm^2 \, s^{-1}}$, hence 
$r_\mathrm{d,hy} \, {\simeq} \,300$~AU, and $J/\mc \, {\simeq} \, 10^{22} \mathrm{\, cm^2 \, s^{-1}}$ so $r_\mathrm{d,hy}\,  {\simeq}\, 400$~AU for run \textsc{SupAS}.
These values roughly meet the binary separation (see Sec.~\ref{sec:diskmass} and Fig.~\ref{fig:t_mrdisk}).
Hence, the binary separation appears to depend on the initial turbulent velocity field.
To gain generality, this experience should be repeated with other realizations of the initial turbulence, but this is beyond the scope of this study.

The circumbinary disk which surrounds the two sink+disk systems is about twice larger than the binary separation and it evolves with time as the two disks interact and as the second disk grows. 
At $t=50$~kyr, which is close to the birth epoch of the two secondary sinks, the circumbinary disk has mostly $\beta \ge 1$ in run \textsc{SupA} (Fig.~\ref{fig:betamaps}).
As shown in Fig.~\ref{fig:circumbeta}, when the secondary sink mass is $5 \Msol$, most of the surrounding gas has become magnetically-supported in run \textsc{SupA}, while thermally-supported gas infall continues actively in run \textsc{SupAS}.
The circumbinary disk does not appear as an isolated structure from the two individual disks, but the fate of this accreting system (individual disks and circumbinary disk) deserves dedicated studies.
In both runs, the binary system is surrounded by a mostly toroidal magnetic field, similarly to unitary systems (runs \textsc{NoTurb} and \textsc{SupA}), as displayed in Fig.~\ref{fig:Bprofiles}. Hence, the magnetic field geometry cannot be used here to discriminate between a unitary and a binary system.

\subsection{What sets the primary disk orientation?}
\label{sec:alignangmom}

\begin{figure*}
\centering
    \includegraphics[width=18cm]{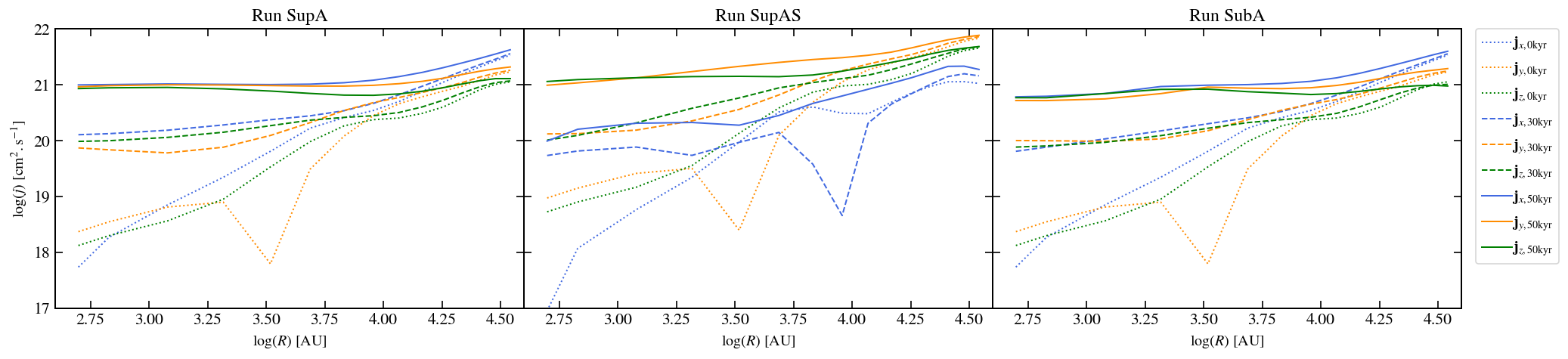}
    \caption{Specific angular momentum $\mathbf{j}=\mathbf{J}/M = \frac{1}{M} \int_{r<R} \mathbf{r} \times \rho \mathbf{v} \, \mathrm{d}V $ as a function of the sphere radius $R$ for runs \textsc{SupA} (left), \textsc{SupAS} (middle) and \textsc{SubA} (right).  Components of the specific angular momentum vector are displayed at $t=0$~kyr (dotted lines), $t=30$~kyr (dashed lines) and $t=50$~kyr (full lines). Time $t=0$~kyr describes the initial conditions, $t=30$~kyr is roughly the first sink formation epoch (a rotating structure is already present), and $t=50$~kyr corresponds to a massive protostar surrounded by its accretion disk. Components along $x$, $y$ and $z$ are shown in blue, orange and green, respectively.}
    \label{fig:dx_j}
\end{figure*}

\begin{figure*}
\centering
    \includegraphics[width=18cm]{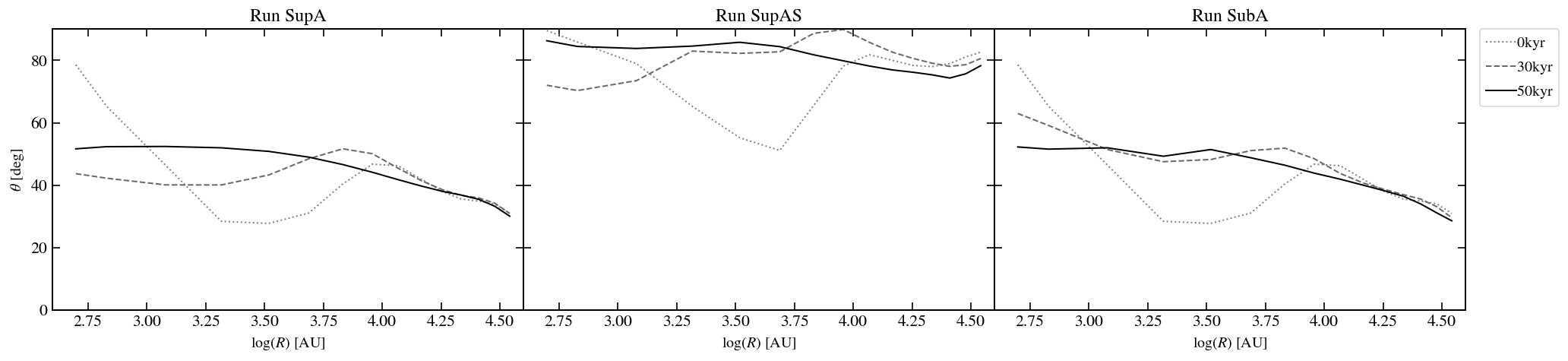}
    \caption{Angle between the specific angular momentum $\mathbf{j}$ and the $x$-axis (the initial magnetic fields orientation) as a function of the sphere radius $R$ for runs \textsc{SupA} (left), \textsc{SupAS} (middle) and \textsc{SubA} (right). The same timesteps as Fig.~\ref{fig:dx_j} are pictured here. The angle is displayed at $t=0$~kyr (dotted lines), $t=30$~kyr (dashed lines) and $t=50$~kyr (full lines)}
    \label{fig:dx_th}
\end{figure*}

One objective of this work is to make progress on the question of whether disks and outflows align with core-scale magnetic fields.
One may expect the disk normal to be preferentially perpendicular to the magnetic field because then the projection of the magnetic braking is smaller and it cannot fully prevent disk formation \citep{joos_protostellar_2012}.
However, this picture may change when turbulence and non-ideal MHD are accounted for. Since both effects individually (see \textit{e.g.} \citealt{joos_influence_2013}, \citealt{hennebelle_magnetically_2016}) and together \citep{lam_disk_2019} solve the magnetic catastrophe, we would expect the disk-magnetic fields orientation to tend towards a random distribution.
Here we investigate the specific angular momentum components, and the alignment between this vector and the large-scale magnetic field (along the $x-$axis) as a function of the Mach number and the magnetic field strength.
The angular momentum vector computed on disk scales ($<10^3$~AU) reveals the disk orientation.

Figure~\ref{fig:dx_j} shows the specific angular momentum $\mathbf{j}$ defined as $\mathbf{j}=\mathbf{J}/M = \frac{1}{M} \int_{r<R} \mathbf{r} \times \rho \mathbf{v} \, \mathrm{d}V $ as a function of the spatial scale $R$ for three epochs: $t=0,30,50$~kyr. We take $r$ as radial vector with respect to the primary sink position, except for the first snapshot, where there is no sink and we take the center of the box. 
We recall that each run has initially a tiny rotational support ($1\%$) of solid-body rotation aligned with the $x-$axis.
In our reference case, run \textsc{NoTurb}, the specific angular momentum is initially aligned with the magnetic field axis and remains so (within less than $6 \deg$, not shown here for readability).
Figure~\ref{fig:dx_j} shows the increase of the local specific angular momentum in the central part of the domain (while the total angular momentum remains constant) as collapse occurs, due to angular momentum transport during the infall.
The angular momentum set by the initial turbulence is dominated by its $y$-component. The dominating component of $\mathbf{j}$ (accounting for both rotation and turbulence) varies with the sphere radius over which it is computed, as a consequence of the initial turbulent velocity field. The initial rotation, aligned with the $x-$axis, dominates at large scales ($>10^4$~AU) in runs \textsc{SupA} and \textsc{SubA} (left and right panels), but not in run \textsc{SupAS} where it is actually smaller. This means that the turbulent gas is in counter-rotation with respect to the initial solid-body rotation imposed.
In the two runs with superalfv\'enic turbulence, \textsc{SupA} and \textsc{SupAS} (left and central panels of Fig.~\ref{fig:dx_j}), the dominating components at disk scales ($<10^3$~AU) at $t=50$~kyr are the initial dominating components at slightly larger scales. These are the $x-$ and $y-$ components in the subsonic run \textsc{SupA} and the $z-$ and $y-$ components in the supersonic run \textsc{SupAS}.
Hence, the disk orientation is set by the initial angular momentum rather than by magnetic fields.

Let us focus on the influence of magnetic fields.
The left and right panels of Fig.~\ref{fig:dx_j} only differ by the magnetic field strength: $\mu=5$ in run \textsc{SupA} (left) and $\mu=2$ in run \textsc{SubA} (right).
At small scales, the component $j_x$ in run \textsc{SupA} is ${\approx}2$ times larger than in run \textsc{SubA}.
This is a consequence of the magnetic braking, \textit{i.e.} the transport of angular momentum outwards. 
It prevents disk formation perpendicularly to the magnetic field and favors configurations where the angular momentum is misaligned with the magnetic field.

In Fig.~\ref{fig:dx_th}, we show the angle between $\mathbf{j}$ (the disk normal) and the $x-$axis, which corresponds to the direction of the large-scale magnetic field.
The orientation of the angular momentum varies significantly with the scale considered and with time.
We get similar results for sub- and supersonic turbulences as \cite{joos_influence_2013}, namely a strong misalignment between $\mathbf{j}$ and $\mathbf{B}$.
The orientation on small scales converges in time as the disk forms and increases in size.
On larger scales, the orientation does not vary except in the most turbulent run, \textsc{SupAS} (Fig.~\ref{fig:dx_th}).
This change is likely due to the velocity field changing configuration, as it becomes dominated by the gravitational pull exerted by the center of mass.
Comparing left and right plots of Fig.~\ref{fig:dx_th} shows that increasing the magnetic field strength, up to the point where the initial turbulence is subalfv\'enic, does not favor the alignment between $\mathbf{j}$ and $\mathbf{B}$.

Overall, the disk normal in our simulations is misaligned ($50-85\deg$, Fig.~\ref{fig:dx_th}) with the large-scale magnetic field, largely because of the initial turbulence.
If the disk formation were a large-scale process, we would expect the disk normal to align with the core-scale angular momentum.
However, as shown in the middle and right panels of Fig.~\ref{fig:dx_j}, $j_y<j_z$ (to be distinguished from $j_x$, which is more affected by magnetic braking) at the disk scale while $j_y>j_z$ at core scales.
The disk orientation here does not appear to be set by the angular momentum direction at core scales because the gas on core scales has not reached yet the center of the domain within our simulated time ($\lesssim 80$~kyr), whereas disk formation occurs within a few $10$~kyr.
This would indicate that disk formation is a "local" process, in agreement with the recent observations in the low-mass regime \citep{gaudel_angular_2020}, but it should first be confirmed for other initial density profiles.
Moreover, disk evolution (and orientation) may be affected by this core scale angular momentum, but longer-time integration is required.

\section{Discussion}
\label{sec:discussion}

\subsection{Comparison with previous works}

As mentioned previously, this work extends the study of C21 to a turbulent medium and with a hybrid radiative transfer method.
In our non-turbulent run \textsc{NoTurb}, we have obtained a final sink mass of $M\,{\simeq}\,15.8 \Msol$.
This value is very similar to what has been found with the FLD method in C21 (their non-ideal MHD run with $\mu=5$ and $E_\mathrm{rot}/E_\mathrm{grav}=1\%$).
The disk radius obtained in run \textsc{NoTurb} also compares well with C21, with ${\sim}100$~AU.
 It shows that, in a magnetized environment, the radiative feedback method does not seem to regulate the sink mass nor the disk radius, up to $M\,{\simeq}\,15.8 \Msol$. 
 Meanwhile, the influence of magnetic fields and ambipolar diffusion is undeniable.
With high-resolution simulations including ohmic dissipation (and no ambipolar diffusion), \cite{kolligan_jets_2018} obtain disks of up to $10^3$~AU. 
While the Hall effect is expected to reduce or increase the disk size depending on the alignment between the rotation axis and the magnetic field axis, this suggests the strong regulating role of ambipolar diffusion \citep{hennebelle_magnetically_2016} to be unique among non-ideal MHD effects.
Taking advantage of this role, we propose that a plasma beta $\beta>1$ (thermal pressure exceeding magnetic pressure) is a good indicator for distinguishing individual disks in the vicinity of protostars.

We report the presence of accretion streamers, which are reminiscent of the accretion channels found by \cite{seifried_accretion_2015} and the bridges in the study of (\citealt{kuffmeier_bridge:_2019}, \citealt{kuffmeier_linear_2020}).
In agreement with \cite{seifried_accretion_2015}, ram pressure dominates over magnetic pressure, but we further show that magnetic pressure is also dominated by thermal pressure.
Those structures develop even in the non-turbulent run \textsc{NoTurb}, and they seem to be associated with magnetic fields effects rather than pure consequences from turbulence, as shown in \cite{kuffmeier_bridge:_2019}.
We note though the the width of the accretion streamers seems to depend on the the initial turbulent level, the larger the turbulence, the wider the streamer. This effect could be a consequence of magnetic turbulent reconnection occurring in the envelop which increases with the turbulent level (\textit{e.g.}, \citealt{santos-lima_disc_2013}). The physical origin of these accretion streamer should be investigated in dedicated studies. 

The present work confirms that disk formation does not occur preferentially perpendicularly to the core-scale magnetic fields, but its orientation is likely driven by turbulence (even subalfv\'enic).
This is in agreement with the study of \cite{machida_misalignment_2019} in the low-mass regime, in which they vary the angle between the rotation axis and the magnetic field direction. They observe that the disk plane is mainly set by their initial rotation (\textit{i.e.} specific angular momentum) axis, even with an initially strong magnetic field ($\mu=1.2$). 

Very few works on massive star formation have reported the formation of a binary system.
Here, we obtained mass ratios of ${\approx} 1.1-1.6$ between the two sinks.
Such balanced mass ratios (of the order of unity) have been obtained in the radiation-hydrodynamical simulations of \cite{krumholz_formation_2009} where they could be integrated for longer times, with final masses of $41.5\Msol$ and $29.2\Msol$ and separation of $1590$~AU.
For comparison, binary separation is $350-600$~AU in run \textsc{SupA} and $400-700$~AU in run \textsc{SupAS} (the orbits are elliptic and the separation is slightly increasing with time).
The studies of \cite{meyer_forming_2018}, focused on the formation of spectroscopic binaries, and \cite{oliva_modeling_2020} on disk fragmentation, are hardly comparable with the present work since they do only use a sink particle algorithm for the primary star.

\subsection{Comparison with observations}

Let us first compare our findings in terms of disk radius with observations.
The disk radius of HH80-81, estimated to be ${\sim}291$~AU \citep{girart_resolving_2018} or Orion Source 1 with $\lesssim 100$~AU \citep{matthews_feature_2010}, are in better agreement with the magnetically-regulated indivual disks radii we obtain than with purely hydrodynamical disks (see \textit{e.g.}, \citealt{kuiper_three-dimensional_2011}, \citealt{mignon-risse_new_2020}).
This is also in line with the upper-limit of $125$~AU set on the disk of S255IR SMA1 \citep{liu_alma_2020}.
Meanwhile, the binary separation is linked to the centrifugal radius that can be derived from the initial turbulent field.
IRAS 16547-4247 is a rare and recent case of massive protostars binary.
Recent measurements with ALMA indicate the presence of jets, a circumbinary disk of radius ${\sim} \, 2500$~AU and individual disks on scale ${\sim} 100$~AU and binary separation of $300$~AU \citep{tanaka_salt_2020}. 
These order-of-magnitude estimates are consistent with runs \textsc{SupA} and \textsc{SubA}, except that they observe hints of counter-rotating disks, indicating preferentially another origin than disk fragmentation.

Furthermore, \cite{aizawa_search_2020} have studied the disk orientation in five nearby star-forming regions and observe a random orientation, consistent with the disk plane being set by turbulence. Hence, it agrees with the present work.
This aspect is also a preliminary step to identify the orientation of outflows (naively expected to be perpendicular to the disk), which has been observationally investigated (\textit{e.g.}, \citealt{hull_understanding_2019}). 

\subsection{Limitations}

Let us now discuss some of the limitations of our approach.
As many other studies of massive star formation (\textit{e.g.}, \citealt{krumholz_formation_2009}, \citealt{kuiper_circumventing_2010}, \citealt{commercon_collapse_2011}), we have chosen a high-mass pre-stellar core for our initial conditions, consistent with the Turbulent Core model of \cite{mckee_formation_2003}.
Indeed, this approach allows us to reach a finest resolution of $5$~AU to capture the disk and the physical scales of interest (the Jeans length and the disk scale height) except very close to the sink, which is rarely done in large-scale simulations without zoom-in procedures.
This condition is strengthened by the non-ideal MHD frame leading to smaller disks than in the hydrodynamical case (see \textit{e.g.}, \citealt{hennebelle_magnetically_2016}, \citealt{masson_ambipolar_2016}).
As claimed by various models, such as the Global Hierarchical Collapse model \citep{vazquez-semadeni_high-_2009} and the Inertial-Inflow model \citep{padoan_origin_2019}, and supported by various observations \citep{schneider_dynamic_2010} the large-scale dynamics is likely playing in a major role in the formation of massive stars, and the isolation we have assumed may be an oversimplification, in particular for the turbulence as it is modelled in this paper.
We leave this to further work.

Protostellar heating could suppress disk fragmentation - or promote it in the shielded disk midplane, because heating increases the Jeans length and stabilizes the gas against gravitational collapse. However, most of the protostellar radiation is absorbed at the disk inner edge and does not heat directly the gas located further way. Very high resolution is required to resolve the photon mean free path, get the exact temperature structure and conclude on the effect of protostellar heating on disk fragmentation. In fact, a cell of gas density $10^{-11} \mathrm{g \, cm^{-3}}$ and opacity to ultraviolet radiation ${\sim}10^2 \mathrm{cm^2 \, g^{-1}}$ would require a spatial resolution of less than $10^{-3}$~AU. Let us note though that it highly depends on the opacity, hence on the source frequency and on the dust density.

Our sink accretion criterion relies on the local Jeans density (as in C21).
Investigating the influence of the accretion method (a density threshold for instance) would be an asset, but it is beyond the scope of this paper.

We enforce gas-radiation decoupling within the sink volume for outflow physics purposes (see Paper II).
Meanwhile, the star exerts a stronger radiative force onto the gas at the first absorption region, \textit{i.e.} at the sink accretion radius (${\sim}20$~AU).
This possibly shifts the disk inner edge (\textit{i.e.} the edge at $20$~AU receives a direct stellar radiative force, perturbing the gravitation-centrifugal equilibrium; see Appendix~\ref{sec:app_fldhy}).

Finally, we use a constant dust-to-gas ratio ($1\%$) within the simulated volume throughout this paper.
Nonetheless, dust is the main contributor to the medium opacity.
Grain growth and sedimentation are expected to occur, affecting this dust-to-gas ratio and therefore the opacities which couple the dust-gas mixture and radiation.
Furthermore, dust grains are charge carriers, so the ionization degree would vary and the non-ideal MHD resistivities together with it.
Hence, dust dynamics should be integrated in collapse calculations (\citealp{lebreuilly_small_2019}, \citealt{lebreuilly_protostellar_2020}).
This would allow one to obtain a dust size distribution that varies dynamically, affecting the dust-and-gas mixture temperature and the effect of magnetic fields which, in turn, sets the disk radial equilibrium.


\section{Conclusions}
\label{sec:ccl}

We have conducted four numerical simulations of massive pre-stellar core collapse including ambipolar diffusion and a hybrid radiative transfer method.
It leads to the formation of thermal pressure-dominated disks, rather than magnetic pressure dominated disks.
We have included an initial velocity field consistent with turbulence, and varied the Mach and Alfv\'enic Mach numbers to consider four runs with respectively, no turbulence (\textsc{NoTurb}), superalfv\'enic-subsonic turbulence (\textsc{SupA}), superalfv\'enic-supersonic turbulence (\textsc{SupAS}) and subalfv\'enic-subsonic turbulence (\textsc{SubA}).
We summarize our results as follows:
\begin{enumerate}
\item Even in the absence of turbulence, asymmetries naturally arise\footnote{Regarding the streamers, the seed is likely numerical in run \textsc{NoTurb} but not in the non-axisymmetric runs \textsc{SupA}, \textsc{SupAS} and \textsc{SubA}, where they are present too.}  via the presence of streamers (thermally-dominated filaments slightly denser than their surroundings) which connect onto the disk off the disk plane, and via the interchange instability which redistributes magnetic flux at the disk edge (see Appendix~\ref{sec:app:intinstab}), breaking the axisymmetry.
\item Keplerian disks form in all runs. They have typical radii of $100-200$~AU around individual stars and are consistent with magnetic regulation. In the superalfv\'enic runs, they are located within a larger rotating structure (circumbinary disk, see below).
In this case, the rotation profile is close to Keplerian rotation within a few hundred AU.
\item We report the formation of stable binary systems when turbulence is superalfv\'enic. They form from disk fragmentation at the extremity of spiral arms rather than initial (core) fragmentation, and follow spiral arm collision. Their binary separation is between $300$~AU and $700$~AU and may be linked to the initial angular momentum (\textit{i.e.} amount of rotation) carried by the turbulent velocity field.
\item We have assessed the misalignment between the disks and core-scale magnetic fields.
Disks orientation appears to be set by the initial angular momentum at scales $\lesssim 10^4$~AU only, in agreement with the previous numerical study of \cite{machida_misalignment_2019}. Meanwhile, the streamers are located in a plane perpendicular to the magnetic field in all runs, but do not influence the disk formation process.
\item A plasma beta (ratio between the thermal pressure and magnetic pressure) larger than one points at structures of interest such as the individual disks and infalling streamers.
\end{enumerate}


This work presents disk accretion as the only accretion mechanism for massive protostars, in opposition to alternative mechanisms such as filamentary accretion or stellar mergers \citep{bonnell_competitive_2001}. The case of radiative Rayleigh-Taylor instabilities is not adressed here because of the debate on the required resolution to get them, but we refer the reader to \cite{krumholz_formation_2009}, \cite{kuiper_stability_2012}, \cite{rosen_unstable_2016} and \cite{mignon-risse_new_2020}.


Our work confirms that multiplicity may be linked to the medium turbulence.
Depending on the models, one may expect this turbulence to be higher in massive star-forming regions (due to radiative outflows and photoionization from other stars, and inflow from large scales), ending up in a higher stellar multiplicity than for low-mass stars.
\\



\begin{acknowledgements}
      This work was supported by the CNRS "Programme National de Physique Stellaire" (\emph{PNPS}). The numerical simulations presented here were run on the \emph{CEA} machine \emph{Alfvén} and using HPC resources from GENCI-CINES (Grant A0080407247). The visualisation of \ramses{} data has been executed with the \href{https://github.com/nvaytet/osyris}{OSYRIS} python package. RMR particularly thanks T. Foglizzo for his insights on the interchange instability.
\end{acknowledgements}

\bibliographystyle{aa} 
\bibliography{Zotero} 

\begin{appendix}
\section{Luminosity injection in the sink particle volume: disk size}
\label{sec:app_fldhy}
 
In this appendix, we investigate the impact of the radiative transfer method and of the kernel function to deposit the luminosity within the sink volume on the disk size.
This is mainly motivated by the need to properly model photon escape within the sink volume to study radiative outflows (see Paper II).

The simulations are similar to run \textsc{NoTurb} presented in the main text, hence they include non-ideal MHD (ambipolar diffusion) but no turbulence.
We run four simulations: two with the Flux-Limited Diffusion ("FLD"), two with the hybrid radiative transfer approach ("HY").
For each method, we test two injection kernels: either the luminosity is deposited uniformly over the sink volume ("uniform"), or only over the central oct ("peaked").

\begin{figure}
\centering
    \includegraphics[width=8cm]{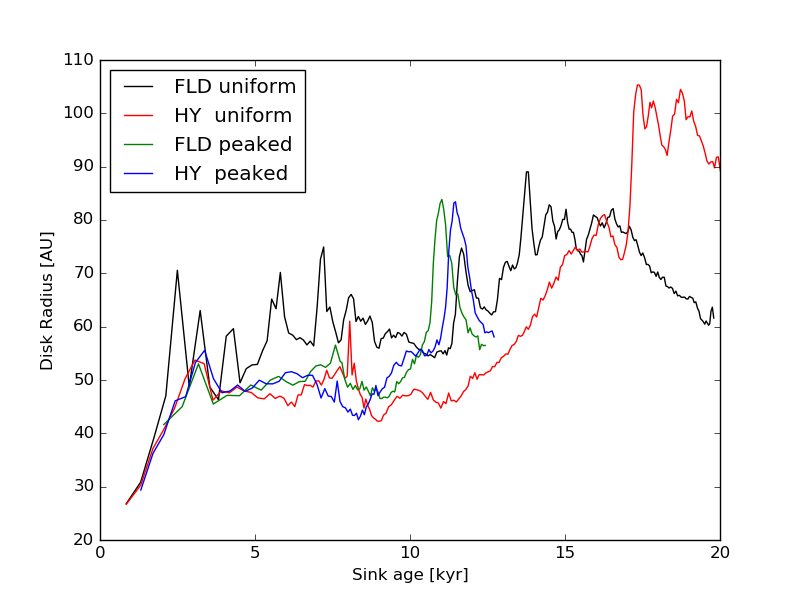}
    \caption{Disk radius as a function of the sink age, for the four runs.}
    \label{fig:app_fldhydisk}
\end{figure}

As shown in Fig.~\ref{fig:app_fldhydisk}, we obtain similar disk sizes in all runs, except in run HY uniform where the disk radius is larger by ${\sim}20$~AU.
This corresponds to the sink accretion (and luminosity injection) radius.
When the luminosity is deposited uniformly up to a radius of $20$~AU, this leads to an additional repulsive force (the direct stellar radiative force) exerted onto the gas over a radius equal to the sink luminosity injection radius, which is $20$~AU.
Hence, a uniform luminosity injection with the hybrid method likely shifts the disk towards a slightly larger radius.

Nevertheless, a uniform injection of luminosity within the sink volume is not physically satisfying.
In fact, the M1 radiative flux which powers the radiative force indirectly depends on the local radiative energy gradient.
If the injection is uniform over the sink volume, radiative energy is more absorbed in the central cells (which sit onto dense gas) than above and below the disk plane (where lower-density gas is located).
This results in a radiative flux oriented towards the central cells and consequently in a spurious radiative force oriented towards the central cells, from above and below the disk plane.
For this reason, we do not adopt a uniform luminosity injection function in this paper but rather set the sink volume as entirely optically-thin.

\section{Interchange instability}
\label{sec:app:intinstab}

\begin{figure*}
\centering
    \includegraphics[width=8cm]{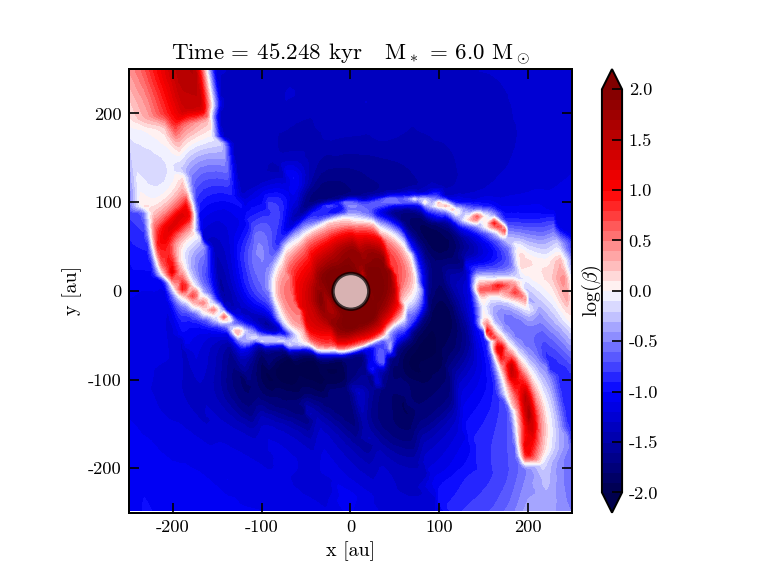}
    \includegraphics[width=8cm]{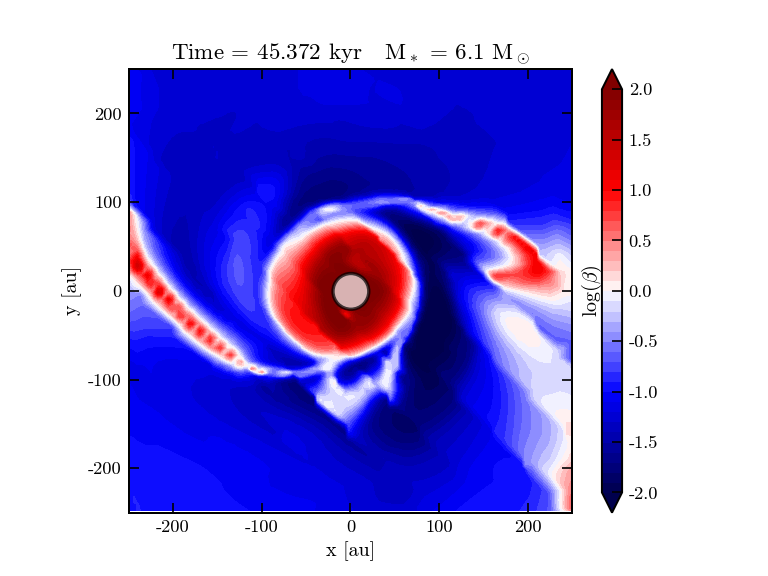}
    \includegraphics[width=10cm]{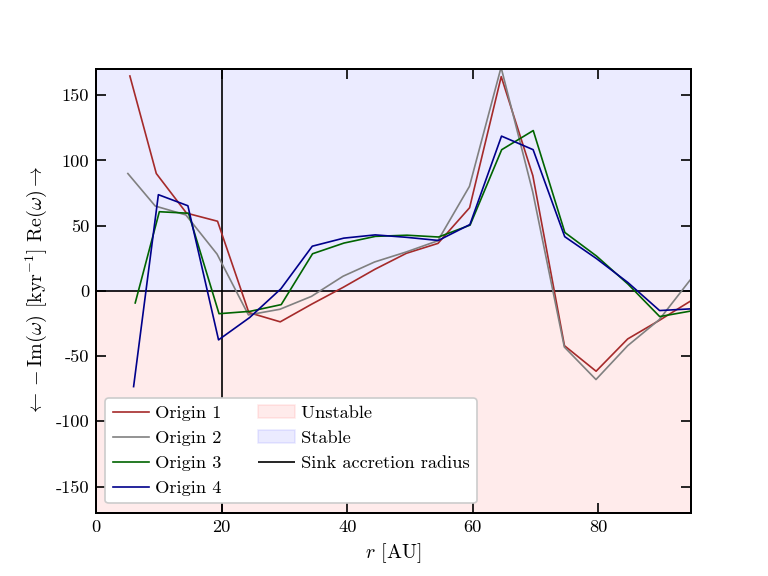}
    \caption{Interchange instability in run \textsc{NoTurb}. Top-left and top-right panels: plasma $\beta$\ at $t=45.25$~kyr (before the instability develops) and $t=45.37$~kyr. Bottom panel: square root $\omega$ of $N^2$ (see Eq.~\ref{eq:instab_growth}): $\mathrm{Im}(\omega)$ gives the growth rate of the interchange instability. We compute it in the $y-$ direction in the disk plane, at $t=45.25$~kyr, taking as origin the four closest points to the sink center.}
    \label{fig:beta_interch}
\end{figure*}

Figure~\ref{fig:beta_interch} shows that a pocket of magnetized plasma is released from the disk edge.
This occurs several times in the simulations but is more difficult to distinguish in the turbulent runs.
In this section we check whether the interchange instability (also called magnetic Rayleigh-Taylor instability), which is a convective instability that redistributes the magnetic flux, is responsible for this.

The instability occurs in the $y-$direction if (see \textit{e.g.} \citealt{lovelace_extragalactic_1981}, \citealp{kaisig_magnetic_1992})
\begin{equation}
\frac{ \norm{ \diffp{((\gamma-1)P+B_\mathrm{x}^2/2)}{y}} }{ \left((\gamma-1)c_\mathrm{s}^2 + v_\mathrm{A}^2  \right)\norm{ \diffp{\rho}{y}}} > 1,
\label{eq:instab_crit}
\end{equation}
where $x$ is the normal direction to the disk, and $v_\mathrm{A}= B/\sqrt{\rho}$ is the Alfv\'en velocity.
The condition of instability is roughly given by the balance between the density gradient set by gravity and the (total) pressure gradient.
We derive the growth rate $\omega=\mathrm{Im}(N)$ analogously to the Brunt-V\"ais\"ala frequency (which is a frequency associated to convective instabilities) from
\begin{equation}
N^2 =\frac{1}{1+\alpha} \left(\frac{\gamma - 1}{\gamma} \diffp{s}{y} + \alpha \diffp{\log(B/\rho)}{y} \right) g_\mathrm{eff},
\label{eq:instab_growth}
\end{equation}
where we defined $\alpha \equiv v_\mathrm{A}^2/c_\mathrm{s}^2$, $s = \frac{1}{\gamma -1} \ln \left(P \, \rho^{-\gamma} \right)$ is the normalized gas entropy, $g_\mathrm{eff} = g - v_\phi^2/r$ is the effective gravity at radius $r$ (sum of the gravitational and centrifugal accelerations).

Bottom panel of Fig.~\ref{fig:beta_interch} shows the square root of $N^2$ in the $y$ direction in the disk plane, varying the $x$ and $z$ coordinates of the origin among the four closest cells to the sink center.
Zones where this value is pure imaginary are unstable, which correspond to the disk edge in multiple directions.
The growth rate at the disk edge is $\omega \, {\approx}\, 70 \mathrm{\, kyr^{-1}}$ for the origin cells of coordinates $[3,2]$~AU (Origin 1) and $[-2,2]$~AU (Origin 2), so the timescale for the instability to develop is $\tau_\mathrm{instab}\, {\approx}\, 14$~yr.
The unstable zone is ${\approx}\,20$~AU wide, in which the gas is flowing at a radial velocity $v_r {\approx}\, 0.8  \mathrm{\,km \, s^{-1}}$ so the advection timescale is $\tau_\mathrm{adv}\, {\approx}\,120$~yr.
Since $\tau_\mathrm{instab}\, \lesssim\, \tau_\mathrm{adv}/3$ \citep{foglizzo_neutrinodriven_2006}, this is consistent with the interchange instability being at work.
When taking the cells $[3,-3]$ (Origin 3) and $[-2,-3]$~AU (Origin 4) as the origin of the profile, it is less clear whether this part of the disk edge is stable or not.
Hence, the small unstable part of the disk edge may explain why this instability is less visible than in \cite{krasnopolsky_protostellar_2012} and too faint to be observable, even though the interchange instability is a good candidate.

\end{appendix}

\end{document}